\documentclass[prb,aps,amsmath,amssymb,amsfonts,floatfix,superscriptaddress,
showpacs,twocolumn]{revtex4}
\usepackage{graphicx}
\usepackage{epsfig}

\def\beq{\begin{equation}}
\def\eeq{\end{equation}}
\def\be{\begin{equation}}
\def\ee{\end{equation}}
\def\iomn{i\omega_n}

\def\cG0{{\cal G}_0}
\def\cG{{\cal G}}

\def\tp{\widetilde{p}}
\def\tphi{\widetilde{\varphi}}
\def\tPsi{\widetilde{\Psi}}

\def\bra{\langle}
\def\ket{\rangle}

\def\vk{{\bf k}}

\def\vr{{\bf r}}
\def\vb{{\bf b}}

\def\br{{\bf r}}

\def\bR{{\bf R}}
\def\bG{{\bf G}}
\def\bT{{\bf T}}

\def\a{\alpha}

\def\g{\gamma}

\def\hP{\hat{P}}

\def\psiG{\psi^{\bG}_{\bk\nu}}
\def\ketkG{|\bk+\bG\rangle}
\def\phik{|\phi^{\g l}_{\bk m}\rangle}
\newcommand{\bk}{\mathbf{k}}

\newcommand{\Bkap}{B_{{\bf k}\alpha'}}
\newcommand{\Bka}{B_{{\bf k}\alpha}}

\newcommand{\bH}{\mathbf{H}}

\newcommand{\Psiknu}{|\Psi_{\bk\nu}\rangle}

\newcommand{\chikm}{|\chi^{\bR}_{\bk m}\rangle}
\newcommand{\tchikm}{|\tilde{\chi}^{\bR}_{\bk m}\rangle}
\newcommand{\chirm}{|\chi^{\bR}_{m}\rangle}
\newcommand{\chitrm}{|\chi^{\bR}_{\bT m}\rangle}
\newcommand{\wfkm}{|w^{\bR}_{\bk m}\rangle}
\newcommand{\wtkm}{|w^{\bR}_{\bT m}\rangle}
\begin{document}

\title{Plane-wave based electronic structure calculations for correlated materials\\
using dynamical mean-field theory
and projected local orbitals}

\author{B.~Amadon}
\affiliation{CEA,
D{\'e}partement de Physique Th{\'e}orique et Appliqu{\'e}e,
Bruy{\`e}res-le-Ch{\^a}tel,
91297 Arpajon Cedex, France}

\author{F.~Lechermann}
\affiliation{I. Institut f{\"u}r Theoretische Physik, Universit{\"a}t Hamburg,
D-20355 Hamburg, Germany}

\author{A.~Georges}
\affiliation{Centre de Physique Th\'eorique, \'Ecole Polytechnique,
91128 Palaiseau Cedex, France}

\author{F.~Jollet}
\affiliation{CEA,
D{\'e}partement de Physique Th{\'e}orique et Appliqu{\'e}e,
Bruy{\`e}res-le-Ch{\^a}tel,
91297 Arpajon Cedex, France}

\author{T.~O.~Wehling}
\affiliation{I. Institut f{\"u}r Theoretische Physik, Universit{\"a}t Hamburg,
D-20355 Hamburg, Germany}

\author{A.~I.~Lichtenstein}
\affiliation{I. Institut f{\"u}r Theoretische Physik, Universit{\"a}t Hamburg,
D-20355 Hamburg, Germany}

\begin{abstract}
The description of realistic strongly correlated systems has recently advanced
through the combination of density functional theory in the local density
approximation (LDA) and dynamical mean field theory (DMFT). This LDA+DMFT method is
able to treat both strongly correlated insulators and metals. Several interfaces
between LDA and DMFT have been used, such as (N-th order) Linear Muffin Tin Orbitals
or Maximally localized Wannier Functions. Such schemes are however either complex in
use or additional simplifications are often performed (i.e., the atomic sphere
approximation). We present an alternative implementation of LDA+DMFT, which keeps the
precision of the Wannier implementation, but which is lighter.
It relies on the
projection of localized orbitals onto a restricted set of Kohn-Sham
states to define the correlated subspace. The method is implemented within the
Projector Augmented Wave (PAW) and within the
Mixed Basis Pseudopotential (MBPP) frameworks. This opens the way to electronic
structure calculations within LDA+DMFT for more complex structures with the
precision of an all-electron method. We present an application to two correlated
systems, namely SrVO$_3$ and $\beta$-NiS (a charge-transfer material),
including ligand states in the basis-set.
The results are compared to calculations done with Maximally Localized Wannier
functions, and the physical features appearing in the orbitally resolved spectral
 functions are discussed.
\end{abstract}


\maketitle

\section{Introduction}

The description of strong electronic correlations in a realistic framework has
become an issue of major importance in current condensed matter research. Due to the
fast progress in the preparation of novel materials, especially in effectively
reduced dimensions, and the advances in experimental techniques in order to probe
such systems, providing an adequate theoretical formalism that can handle explicit
many-body effects in a material-specific context is crucial. It is evident that
standard density functional theory (DFT) in the local density approximation (LDA)
cannot meet those demands and, at least, has to ally with manifest
many-body techniques. In this respect, the combination of DFT-LDA with the
Dynamical Mean-Field Theory (DMFT), the so called LDA+DMFT approach, has proven to
be an invaluable method to face the challenge. Already numerous studies within the
LDA+DMFT framework have shown that this theory is capable of describing the
effects of strong correlations in a realistic context, such as Mott transitions and
volume-collapse transitions in $d$ and $f$-electron systems, effective mass enhancement,
local moment formation and magnetism, spectral-weight transfers, orbital physics, etc...

In view of these successes, it is crucial to push further
the range of applicability of LDA+DMFT methods. To this aim, implementing
LDA+DMFT within those electronic structure methods which are highly accurate and
allow for the possible treatment of larger systems, is certainly an important endeavor.
One of the main aims of this paper is to report on such new implementations, within
a projector augmented-wave (PAW) and a mixed-basis pseudopotential (MBPP) method.
The formalism used in this paper is actually quite general and allows for an implementation of
LDA+DMFT in a very large class of electronic structure methods. We present this formalism in an
arbitrary basis set, which should make this task easier for other implementations. However, in
our specific implementation, we make an extensive use of the Bloch basis set of Kohn-Sham orbitals.
In order to implement LDA+DMFT, we use local orbitals constructed by projecting
atomic-like orbitals onto a restricted set of Bloch states, a strategy similar to
that introduced by Anisimov {\sl et al.}~\cite{Anisimov05}.

Another important route is to investigate
how to optimize the application of LDA+DMFT for a
specific material, or class of materials of interest.
Indeed, it is important to realize that, while the results of course do not
depend on the specific basis set used in the calculation, they will actually depend
on the specific set of {\it local orbitals} for which many-body effects will be taken into
account, or more precisely (following the terminology introduced in
[\onlinecite{Lechermann06}]) on the ``correlated'' subspace ${\cal C}$ of the full Hilbert
space spanned by these orbitals. Indeed, the DMFT treatment applies local interaction terms
to those orbitals only, and furthermore neglects all non-local components of the
self-energy. This notion of locality is defined with respect to the specific choice of
the local orbitals defining ${\cal C}$, and the accuracy of the DMFT approximation cannot be
expected to be identical for different choices.

Early implementations~\cite{Anisimov_lda+dmft_1997,Lichtenstein_lda+dmft_1998,Savrasov_Kotliar_pu_nature_2001}
of LDA+DMFT used the linear muffin-tin orbital~\cite{Andersen_lmto_1975_prb} (LMTO)
framework, and the correlated orbitals were frequently identified with a specific
subset of the LMTO basis functions, having $d$- or $f$-character. There is no
specific reason to pick the local orbitals as members of the basis set however, and
a set of atomic-like functions may prove to be better from the physical point of
view.
Recently, several works have used different kinds of Wannier functions (WFs) as
correlated orbitals, starting with the work of
Pavarini {\it et al.}~\cite{Pavarini04} using
Nth-order muffin-tin orbitals~\cite{Andersen_NMTO_00} (NMTO).
Anisimov {\it et al.}~\cite{Anisimov05} used WFs constructed from a
projection on a subset of the Bloch functions~\cite{Cloizeaux63,ku02}, and
Lechermann {\it et al.} used maximally-localized~\cite{Marzari97} WFs and compared
the results to those using NMTOs. 

The energy window spanned by the basis functions which are retained in the implementation, and
the spatial extension of the local orbitals defining ${\cal C}$ are important physical issues
for the description of a given materials. Those issues become particularly important for
charge-transfer materials (e.g. late transition-metal oxides, sulfides or selenides) in which the
ligand states must be kept in the basis set in order to reach a satisfactory physical picture.
In the present article, we address these issues and provide explicit comparisons between
calculations performed with more spatially extended local orbitals (hence spanning a smaller
energy window when projected onto Bloch functions) and more localized local orbitals
(hence spanning a wider energy window).

This article is organized as follows. Sec.~\ref{sec:theory} is devoted to a presentation of
the general theoretical framework. Secs.~\ref{sec:srvo3} and \ref{sec:nis} are devoted to
applications to two compounds, which are considered as tests for the method: the transition-metal
oxide SrVO$_3$ in Sec.~\ref{sec:srvo3} and the charge-transfer sulfide NiS in Sec.~\ref{sec:nis}.

\section{Theoretical framework}
\label{sec:theory}

\subsection{LDA+DMFT formalism in an arbitrary basis set}
\label{se:localorb}

In order to implement DMFT within realistic electronic structure calculations
of correlated-electron materials, one has to set up a formalism which keeps track
of the real-space (i.e., quantum chemical) and the reciprocal-space (i.e., solid-state) aspects
on an equal footing, while being computationally efficient.
Following the work of Lechermann {\it et al.}\cite{Lechermann06}, we therefore
distinguish between the complete basis set $\{|\Bka\rangle\}$ in which the full
electronic structure problem on a lattice is formulated (and accordingly the
lattice Green's function is represented), and local orbitals $\chirm$ which span
a ``correlated'' subspace ${\cal C}$ of the total Hilbert space.
Many-body corrections beyond LDA will be considered inside this subspace.
The index $\a$ labels the basis functions for each wave vector
$\bk$ in the Brillouin zone (BZ). The index $\bR$ denotes the correlated atom within 
the primitive unit cell, around which the local orbital $\chirm$ is centered, and
$m$=$1,\ldots,M$ is an orbital index within the correlated subset.
Projection on that subset, for atom type $\bR$, is done with the projection
operator:
\begin{equation}
\hP^{(\mathcal{C})}_{\bR} \equiv \sum_{m\in\mathcal{C}}
|\chi^{\bR}_{m}\rangle \langle \chi^{\bR}_{m}|\,\,.
\label{proop}
\end{equation}
In Ref.~\onlinecite{Lechermann06}, it was shown that the DMFT self-consistency condition, which
relates the impurity Green's function $\bG^{\bR}_{\rm{imp}}$ to the Green's function
of the solid computed locally on atom $\bR$, reads:
\begin{eqnarray}\label{eq:scc_basis}
G^{\bR,\rm{imp}}_{mm'}(\iomn) &=&
\sum_{\bk}\sum_{\alpha\alpha'}
\langle\chi^{\bR}_{\bk m}|\Bka\rangle\langle\Bkap|
\chi^{\bR}_{\bk m'}\rangle
\times \nonumber \\
&&\hspace{-2cm}\times\left\{\left[\iomn+\mu-\bH_{\rm{KS}}(\bk)-
\mathbf{\Delta\Sigma}(\bk,\iomn)\right]^{-1}\right\}_{\alpha\alpha'}\,,
\end{eqnarray}
In this expression:
\begin{equation}
\chikm=\sum_{\bT}{\rm e}^{i\bk\cdot(\bT+\bR)}\chitrm
\end{equation}
denotes the Bloch transform of the local orbitals whereby
$\bT$ denotes the Bravais lattice translation vectors.
Note that the object in the second line of Eq.~(\ref{eq:scc_basis}) is, of course,
nothing else than the full lattice Green's function
$G_{\a\a'}(\bk,\iomn)$ in the chosen $\{|\Bka\rangle\}$ basis.
The Kohn-Sham (KS) hamiltonian $\bH_{\rm{KS}}(\bk)$ is obtained by solving
the Kohn-Sham equations, which yields eigenvalues $\varepsilon_{\bk\nu}$ and Bloch
wave-functions $|\Psi_{\bk\nu}\rangle$ ($\nu$ is
the band index). It can be expressed in the  $\{|\Bka\rangle\}$ basis set as:
\begin{eqnarray}
\bH_{{\rm KS},\alpha\alpha'}(\bk) &=&
\sum_{\nu}\langle\Bka|\Psi_{\bk\nu}
\rangle\,\varepsilon_{\bk\nu}\,\langle\Psi_{\bk\nu}|\Bkap\rangle\,\,.
\end{eqnarray}
In order to obtain the self-energy for the full solid, one has to promote
(``upfold'')
the DMFT impurity self-energy $\Sigma^{\rm{imp}}_{mm'}$ to the lattice
via~\cite{Lechermann06}:
 \begin{eqnarray}\label{eq:scc_self}
\Delta\Sigma_{\alpha\alpha'}(\bk,\iomn)&=&
\sum_{\bR}\sum_{mm'} \langle\Bka|\chi^{\bR}_{\bk m}\rangle\langle
\chi^{\bR}_{\bk m'}|\Bkap\rangle\times\nonumber\\
&&\hspace{-1cm}\times\left[\Sigma_{mm'}^{\rm{imp}}(\iomn)-
\Sigma_{mm'}^{\rm{dc}}\right]\,,
\end{eqnarray}
whereby a double-counting correction $\Sigma_{mm'}^{\rm{dc}}$, taking care of
correlation effects already accounted for in the LDA hamiltonian (see appendix
\ref{dc-app}) has to be included in the general case.

The above equations form the general LDA+DMFT framework, in a general
arbitrary basis set. Any specific implementation must then make a
choice for:
\begin{itemize}
\item[i)] The set of local orbitals
$\{\chirm\}$ spanning the correlated subspace
\item[ii)] The specific basis set $\{|\Bka\rangle\}$
in which these equations will be implemented
\end{itemize}
It is important to realize that the results will certainly depend on the
specific choice of $\{\chirm\}$: the quality of the DMFT approximation will indeed
depend on how the local orbitals are picked such as to minimize non-local
contributions. In contrast, for a given choice of $\{\chirm\}$'s, the results should
be in principle independent of the basis set $\{|\Bka\rangle\}$ which is chosen for the
implementation. However, in practice, considerations of numerical efficiency do limit
the size of the basis which can be handled. Indeed, Eq.~(\ref{eq:scc_basis}) involves
inverting a matrix (at each $\bk$-point and for each frequency) of size $N_b\times N_b$,
in which $N_b$ is the number of basis functions which are eventually retained. Hence,
in practice, one will restrict the basis set to a certain set ${\cal W}$ of bands, as
will be described in more details below.
Regarding the choice of local orbitals, we recall that, in Ref.~[\onlinecite{Lechermann06}],
different kinds of Wannier functions were used and compared to one another, namely
maximally-localized~\cite{Marzari97} and $N$th-order linear-muffin-tin orbitals
(NMTO)~\cite{Andersen_NMTO_00}. The construction of such functions requires rather
sophisticated procedures.
In the present work we use a somewhat lighter implementation,
with the same demands on accuracy, by constructing the $\{\chirm\}$'s out of
entities that are already existing in most of the common band-structure codes,
namely using the decomposition of local atomic-like orbitals onto the basis function
retained in the set ${\cal W}$. This is very similar in spirit to the
construction proposed by Anisimov {\sl et al.}~\cite{Anisimov05} in the LMTO
framework. The present construction, and the basic equations developed in this section,
are fairly general however and can be implemented in an arbitrary electronic
structure code.

\subsection{LDA+DMFT formalism in the Bloch basis set}

In this section, we focus on a very simple choice for the basis set, namely
the Bloch basis itself $\{|\Bka\rangle\}$$=$$\{|\Psi_{\bk \nu}\rangle\}$.
This basis is most conveniently used, since it is a direct output of any DFT-LDA
calculation and furthermore diagonalizes the KS hamiltonian
$\bH^{\rm KS}_{\nu\nu'}(\bk)= \delta_{\nu\nu'}\varepsilon_{\bk\nu}$.

The basic LDA+DMFT equations in the Bloch basis set are easily written up, using the projection
matrix elements of the local orbitals onto the Bloch functions, defined as:
\begin{equation}\label{eq:pro_chipsi}
P^{\bR}_{m\nu}(\bk)\equiv\langle\chi^{\bR}_{\bk m}|\Psi_{\bk\nu}\rangle\,\,,\quad
P^{\bR *}_{\nu m}(\bk)\equiv\langle\Psi_{\bk\nu}|\chi^{\bR}_{\bk m}\rangle\,\,.
\end{equation}
Equations (\ref{eq:scc_basis}),(\ref{eq:scc_self}) then read:
\begin{eqnarray}
G^{\bR,{\rm imp}}_{mm'}(\iomn) &=&
\sum_{\bk,\nu\nu'}
P^{\bR}_{m\nu}(\bk)\,G^{\rm bl}_{\nu\nu'}(\bk,\iomn)\,
P^{\bR*}_{\nu' m'}(\bk)\,\,\,\label{eq:g_allband}\\
\Delta\Sigma^{\rm bl}_{\nu\nu'}(\bk,\iomn)&=&
\sum_{\bR}\sum_{mm'}P^{\bR*}_{\nu m}(\bk) \,\Delta\Sigma^{\rm imp}_{mm'}(\iomn)
\,P^{\bR}_{m'\nu'}(\bk)\,\label{eq:sig_allband}
\end{eqnarray}
where
\begin{eqnarray}
&&G^{\rm bl}_{\nu\nu'}(\iomn)=\left\{\left
[(\iomn+\mu-\varepsilon_{\bk\nu})\delta_{\nu\nu'}-
\mathbf{\Delta\Sigma}_{\rm bl}(\bk,\iomn)\right]^{-1}\right\}_{\nu\nu'} \\
&&\Delta\Sigma^{\rm imp}_{mm'}(\iomn)=\Sigma_{mm'}^{\rm{imp}}(\iomn)-
\Sigma_{mm'}^{\rm{dc}}\quad.
\end{eqnarray}

\subsection{Truncating the Bloch basis set, and choice of local orbitals}

As pointed out above, it is computationally impossible to implement these LDA+DMFT equations
without restricting oneself to a finite subspace of $N_b$ Bloch functions. Those
Bloch functions span a certain energy window, corresponding to a subspace ${\cal W}$ of the
total Hilbert space.
Naturally, the local atomic-like orbitals $\chikm$ will in general have a decomposition
involving {\it all} Bloch bands. The Bloch-transform of these local orbitals read:
\begin{equation}\label{eq:Bloch_dec_full}
\chikm=\sum_{\nu}\langle\Psi_{\bk\nu}|\chi^{\bR}_{\bk m}\rangle
|\Psi_{\bk\nu}\rangle\,\,.
\end{equation}
Note that, starting from an orthonormalized set of local orbitals
$\langle\chi_{m\bk}^\bR|\chi_{m'\bk'}^{\bR'}\rangle=\delta_{mm'}\delta_{\bR\bR'}\delta_{\bk\bk'}$,
it is easily checked
that the matrix $\langle\Psi_{\bk\nu}|\chi^{\bR}_{\bk m}\rangle$ is unitary
(from the completeness of the Bloch basis). Hence
the $\chi_m$'s can formally be viewed as Wannier functions associated with the complete basis set
of all Bloch states.
This property no longer holds, however, when the sum in (\ref{eq:Bloch_dec_full}) is
restricted to the subset ${\cal W}$ of Bloch band. Defining:
\begin{equation}
\tchikm\equiv\sum_{\nu\in {\cal W}}\langle\Psi_{\bk\nu}|\chi^{\bR}_{\bk m}\rangle
|\Psi_{\bk\nu}\rangle
\end{equation}
it is seen that the functions $\tchikm$ are not true Wannier functions associated with the
subspace ${\cal W}$ since the truncated projection matrix is no longer unitary. However, these
functions can be promoted to true Wannier functions $\wfkm$ by
orthonormalizing this set according to:
\begin{equation}
\wfkm=\sum_{\bR'm'} S^{\bR\bR'}_{mm'}(\bk)|\tilde{\chi}^{\bR'}_{\bk m'}\rangle\,\,,
\label{eq:wkfm}
\end{equation}
where ${\bf S}^{\bR\bR'}(\bk)$ is given by the inverse square root of the overlap
matrix between the Wannier-like orbitals, i.e.,
\begin{eqnarray}
O^{\bR\bR'}_{mm'}(\bk)&\equiv&\langle\tilde{\chi}^{\bR}_{\bk m}
|\tilde{\chi}^{\bR'}_{\bk m'}\rangle=
\sum_{\nu\in {\cal W}}P^{\bR}_{m\nu}(\bk)P^{\bR' *}_{\nu m'}(\bk)\\
S^{\bR\bR'}_{mm'}(\bk)&\equiv&
\left\{\left({\bf O}(\bk)\right)^{-1/2}\right\}_{mm'}^{\bR\bR'}
\end{eqnarray}
Naturally, the functions $w_m^\bR$ are {\it more extended} in space than the original atomic-like
functions $\chi_m^\bR$ since they can be decomposed on a {\it smaller} number of
Bloch functions, spanning a restricted energy range.

In the end, LDA+DMFT is implemented by taking for ${\cal C}$ the correlated subset
generated by the set of functions $\wfkm$. Since those functions have a
vanishing overlap with all Bloch functions which do not belong to the set ${\cal W}$, the
LDA+DMFT equations can now be put in a computationally tractable form, involving
only a $N_b\times N_b$ matrix inversion within the selected space ${\cal W}$. Hence,
the equations which are finally implemented read:
\begin{eqnarray}
G^{\bR,{\rm imp}}_{mm'}(\iomn)\hspace{-0.1cm} &=&\hspace{-0.5cm}
\sum_{\bk,(\nu\nu')\in {\cal W}}
\bar{P}^{\bR}_{m\nu}(\bk)\,G^{\rm bl}_{\nu\nu'}(\bk,\iomn)\,
\bar{P}^{\bR*}_{\nu' m'}(\bk)\,,\,\,\label{eq:g_limband}\\
\Delta\Sigma^{\rm bl}_{\nu\nu'}(\bk,\iomn)&=&
\sum_{\bR}\sum_{mm'}\bar{P}^{\bR*}_{\nu m}(\bk) \,\Delta\Sigma^{\rm imp}_{mm'}(\iomn)
\,\bar{P}^{\bR}_{m'\nu'}(\bk)\,,\label{eq:sig_limband}
\end{eqnarray}
where 
\begin{eqnarray}
\bar{P}^{\bR}_{m\nu}(\bk)&\equiv&
\sum_{\bR'm'}S^{\bR\bR'}_{mm'}(\bk)\,P^{\bR'}_{m'\nu}(\bk)\,\,,\\
\bar{P}^{\bR*}_{\nu m}(\bk)&\equiv&
\sum_{\bR'm'}S^{\bR\bR'*}_{m'm}(\bk)\,P^{\bR'*}_{m'\nu'}(\bk)\quad.
\end{eqnarray}
It is important to realize that the truncation to a limited set of Bloch functions
was not reached by simply neglecting matrix elements between the local orbitals and
Bloch functions outside this set, but rather by constructing a new set of (more extended)
local orbitals such that the desired matrix elements automatically vanish, hence
redefining ${\cal C}$ accordingly. In this view, the choice of ${\cal C}$ and
of ${\cal W}$, although independent in principle, become actually inter-related.

We also note that it is not compulsory to insist on forming true Wannier functions
out of the (non-orthogonal) set $\tchikm$.
It is perfectly legitimate formally to choose the correlated
subspace ${\cal C}$ as generated by orbitals having a decomposition in ${\cal W}$, but
not necessarily unitarily related to Bloch functions spanning ${\cal W}$.
Although orthogonality of the $\chi_m$'s is also not compulsory,
several (but not all) impurity solvers used within DMFT do require however that
the $\chi_m's$ be orthogonal on a given atomic site.
One possibility, for example, is to
orthonormalize this set on identical unit-cells only, i.e requiring that $\wtkm$
and $|w^{\bR'}_{\bT'm'}\rangle$ in
real space are orthogonal for $\bT=\bT'$, but not in neighboring cells
$\bT\neq \bT'$. This amounts to orthonormalize the $\tchikm$ set with respect to the
$\bk$-summed overlap matrix, instead of the one computed at each $\bk$-point.

In our actual implementations, the wave-functions spanning the correlated
subspace ${\cal C}$ are obtained by following the above orthonormalization procedure,
starting from atomic-like orbitals $\chi_m^\bR$ centered on the atomic site $\bR$ in the
primitive unit-cell. These local
orbitals are either {\sl all-electron atomic partial waves} in the PAW framework, or
{\sl pseudo-atomic wave functions} when using the MBPP code.
Since, in the present work, we are not dealing with
full charge self-consistency including self-energy effects, the
matrix elements $P_{m\nu}^\bR(\bk)$ and the wave-functions $\wfkm$
can be computed once and for all at the beginning of the DMFT cycle~\cite{Lechermann06}.
Details on the specific construction of the local orbitals used in this article,
and the corresponding calculation of (\ref{eq:pro_chipsi}) are summarized in
appendix~\ref{appendix:psichi}.

\subsection{Physical considerations on the choice of
the correlated subspace ${\cal C}$ and
of the Wannier/Bloch space ${\cal W}$}

Let us now discuss some physical considerations regarding the choice of the
truncated Bloch space ${\cal W}$ when using the LDA+DMFT framework
to describe a given material. Operationally, this means that a certain number of
Bloch bands $N_b$ (spanning a certain energy window) will be retained when
solving Eqs.~(\ref{eq:g_limband},\ref{eq:sig_limband}). As discussed above, the
choice of ${\cal W}$ also influences the actual definition of the correlated subspace
${\cal C}$, since we require that the orbitals generating ${\cal C}$ can be expanded
upon the basis functions generating ${\cal W}$.

Let us consider, to be specific, the case of a transition-metal oxide, such as for
example SrVO$_3$. This material, which is described in more details in the following section,
has a set of  three $t_{2g}$ bands, well separated from both the O$_{2p}$ and
$e_g$ bands, and containing nominally one $d$-electron. We can make two rather extreme
choices when describing this material with LDA+DMFT:

i) Focus only on a very limited set of low-energy Bloch bands, such as the three
$t_{2g}$ bands, and generate ${\cal W}$ just from the three corresponding Bloch
functions. In this case, we shall have also ${\cal C}={\cal W}$, and the
$\wtkm$ will be Wannier functions unitarily related to these three Bloch bands.
Since these bands span a narrow energy window, this also means that these Wannier
functions will be rather extended spatially: although centered on Vanadium atoms, they
will have a sizeable contribution on neighboring oxygen atoms as well. This kind of
approach has been emphasized and studied in details in Ref.~[\onlinecite{Lechermann06}].
Of course, it is then out of the scope of such approaches to investigate the indirect
effects of correlations on bands other than the $t_{2g}$ ones.

ii) Alternatively, one may choose a large energy window to define ${\cal W}$, including
in particular all Bloch bands corresponding to O$_{2p}$, V-$t_{2g}$ and V-$e_g$. Then,
the orbitals $\wtkm$ defining the correlated subspace ${\cal C}$ may be chosen as having a
component on Bloch states spanning a much larger energy range. As such, they will
be more localized spatially, i.e closer to (vanadium) atomic-like orbitals.
When working with such an enlarged space ${\cal W}$, the physics of O$_{2p}$ and
$e_g$ states can also be addressed.

One of the goals of the present paper is to present and compare calculations done
with such different choices of ${\cal W}$ and ${\cal C}$. Of course, in a fully first-principle
approach, the screened interaction matrix elements should also be calculated (e.g in a
GW framework) in a manner which
is consistent with these choices of Hilbert spaces. This is left for future investigations
however, and in the present work, these matrix elements will be taken as parameters.

Let us note that the local orbitals are constructed in the present paper from an atomistic
point of view. Hence there is no ``entangling problem'' as the one encountered when
constructing maximally localized Wannier functions for strongly hybridized band complexes. Of course,
in the present formalism, the Wannier functions obtained by projection of atomic orbitals
onto Bloch states belonging to a narrow energy window are not maximally-localized
in the sense of Ref.~\onlinecite{Marzari97}, but this feature does not bring essential differences
in the results, as clear from the results reported below.

Finally, in order to relate the LDA+DMFT results to experiments performed using
e.g, photoemission spectroscopy, the real-frequency spectral functions must be
obtained. This can be done using e.g. a maximum entropy treatment of Monte Carlo data,
but is important to understand how to connect the calculated quantities to
physically observable spectra. The most direct output of the LDA+DMFT calculation
is the local impurity spectral function, obtained (for orbital $m$) as:
\begin{equation}
A_m^{\rm{imp}}(\omega)\,\equiv\, -\frac{1}{\pi}
\rm{Im} G_{mm}^{\rm{imp}}(\omega+i0^+)
\label{eq:Aimp}
\end{equation}
This also corresponds to matrix elements of the full Green's function of the solid within
atomic-like orbitals $\chi_m$'s. This however, is {\it not} a quantity which can be easily
related to photoemission experiments, since the $\chi_m$'s, when very localized spatially,
extend over a large energy range. When considering a certain energy, contributions of
other electrons with a different orbital character than the $\chi_m$'s will contribute significantly
to the photoemission signal. Instead, if one is interested in the measured spectral function in a
given energy window, one must consider the matrix elements of the full Green's function
within Bloch (or Wannier) functions spanning that energy range, namely:
\begin{equation}
A_\nu (\omega)\,\equiv\, -\frac{1}{\pi}
\rm{Im} \sum_\bk\,G^{\rm{bl}}_{\nu\nu}(\bk,\omega+i0^+)
\label{eq:Abloch}
\end{equation}
This quantity can be obtained either by first reconstructing the local self-energy
on the real-frequency axis by analytical continuation of the impurity Green's function, or
by direct analytic continuation of the imaginary-frequency
Bloch Green's function $\bG^{\rm bl}(\bk,\iomn)$.

\section{Application: SrVO$_3$}
\label{sec:srvo3}

SrVO$_3$ is a $t_{2g}^1e_g^0$ metal. It is a good test case for LDA+DMFT calculations
because it is cubic and non magnetic and also
the $t_{2g}$ bands are isolated from both $e_g$ and oxygen $p$ bands -- in the LDA
bandstructure. Numerous calculations (including LDA+DMFT) and experiments have been
done on this compound
~\cite{imada_mit_review,fujimori_pes_oxides,maiti_2001,maiti_phd,
inoue_casrvo3_1995_prl,sek04,lie03,nek05,pav04,yos05,wad06,sol06,nek06,egu06}.
From theses studies, it appears that
there is a need to include correlations to describe correctly this compound.
This is thus an ideal system to benchmark this new implementation.
\subsection{LDA}
\begin{figure}
{\resizebox{8.3cm}{!}
{\rotatebox{0}{\includegraphics{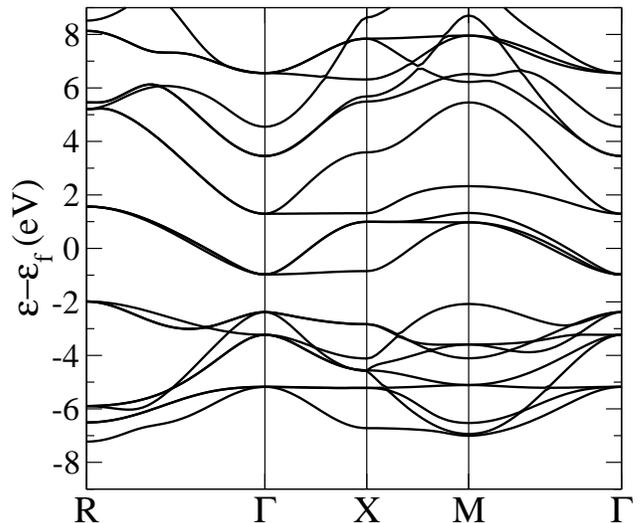}}}}
\caption{LDA band structure for SrVO$_3$.}
\label{fig:bsSrVO3}
\end{figure}

\begin{figure}
{\resizebox{8.3cm}{!}
{\rotatebox{0}{\includegraphics{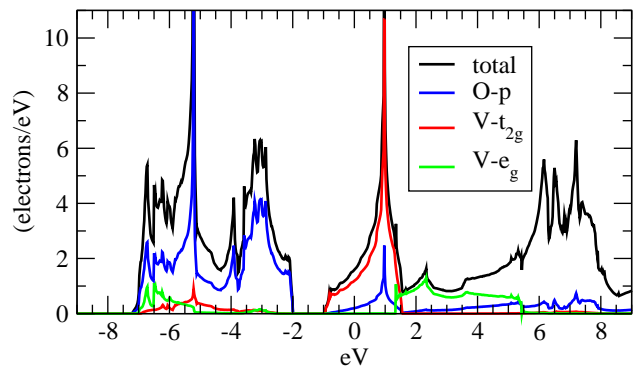}}}}
\caption{(Color online) LDA total and projected density of states for SrVO$_3$.}
\label{fig:pdosSrVO3}
\end{figure}

\begin{figure}
{\resizebox{8.3cm}{!}
{\rotatebox{0}{\includegraphics{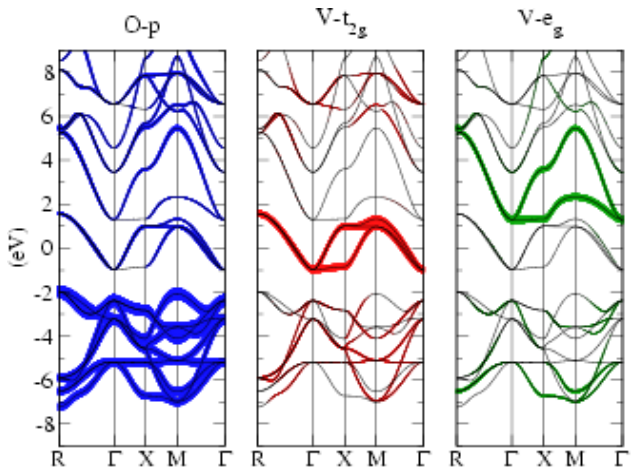}}}}
\caption{(Color online) LDA band structure for SrVO$_3$ computed in PAW,
with "Fatbands" to show the amplitude of the projection of each band on a given atomic orbital
(O-$p$, V-t$_{2g}$ and V-$e_g$).}
\label{fig:fatSrVO3}
\end{figure}

For the PAW calculations, semicore states of V and Sr are treated in the valence.
Valence states for Sr, V and O thus include respectively $4s4p5s$, $3s3p4s3d$ and $2s2p$ states.
PAW matching radius are 1.52 a.u., 1.92 a.u. and 2.35 a.u. respectively. The experimental
cubic crystal structure is used (space groupe \emph{Pm$\bar{3}$m} with lattice constant
of 7.2605 a.u.). Atomic data are generated using ATOMPAW\cite{atompaw,AtompawAbinit}.
Calculations are done with the PAW code ABINIT\cite{abinit1,torrent07}.
The density of states and the LDA bandstructure are shown on Fig. \ref{fig:bsSrVO3} and \ref{fig:pdosSrVO3}.
The projection of the density of states on O-$p$, V-$t_2g$ and V-$e_g$  and the character of the bands
(see Fig. \ref{fig:fatSrVO3}) show that bands with Oxygen $p$ and with Vanadium $t_2g$ characters  are indeed
isolated from the others.  The hybridization between Oxygen and Vanadium orbitals is nevertheless clearly seen.


\subsection{LDA+DMFT}
In this section, we will present the results of our new scheme
based on the projection of Bloch states upon Local Orbitals (Projection
on Local Orbitals (PLO)). As previously emphasized, the number of  KS bands
used for the projection have to be chosen in a given range of energy. The extension of
the Wannier-like renormalized orbitals (Eq. \ref{eq:wkfm}) will depend on this choice.
We will use the fact that the band structure of SrVO$_3$ is made
of isolated blocs. The choices of $\mathcal{W}$ are summarized in table \ref{tab:WN}.
We will compare the results with LDA+DMFT
calculations done~\cite{Lechermann06} with Maximally
Localized Wannier Functions (MLWF) (see appendix \ref{app:Wannierpaw}).

\begin{table}[h]
\begin{tabular}{lcccccccc}
\\\hline\hline
\hline
 $N_{b}$ & $N_{\rm imp}$  & Energy range (eV)\\
\hline\hline
    3      &   3    &  -1.5$\rightarrow$1.8\\
    12     &   3    &  -8.0$\rightarrow$1.8\\
    21     &   5    &  -8.0$\rightarrow$13.0\\
\hline\hline
\label{tab:WN}
\end{tabular}
\caption{(Color online) Different choices for the number of KS states $N_{b}$ used within the PLO scheme of
LDA+DMFT. The energy range spanned by these bands is also indicated. $N_b=3$ corresponds to only the three bands 
of mainly t$_{2g}$ character. $N_b=12$ corresponds to the 9 Oxygen-$p$ bands and the 3 Vanadium-$t_{2g}$ bands (see
Fig. \ref{fig:fatSrVO3}).
$N_b=21$ corresponds to a large number of bands, including $e_{g}$ bands and 7 bands above.
$N_{\rm imp}$ is the number of orbitals in the impurity model. 
When used, MLWF will be extracted with the same parameters.}
\label{tab:NW}
\end{table}
Calculations have been carried out at T=0.1 eV.
A density-density interaction vertex is used \cite{Fresard97}, with J=0.65 eV,
similarly to another study\cite{Lechermann06}.
The Hubbard parameters U=4 eV and U=6 eV have been used
in the calculations. The impurity problem of DMFT is solved with Hirsch-Fye Quantum Monte Carlo with 128 time slices.
The Around Mean Field formulation of the double counting is used in these calculations (see appendix \ref{dc-app}).

\subsubsection{$t_{2g}$ bands in the basis for $\mathcal{W}$ ($N_{b}=3$).}
\begin{figure}
{\resizebox{8.3cm}{!}
{\rotatebox{0}{\includegraphics{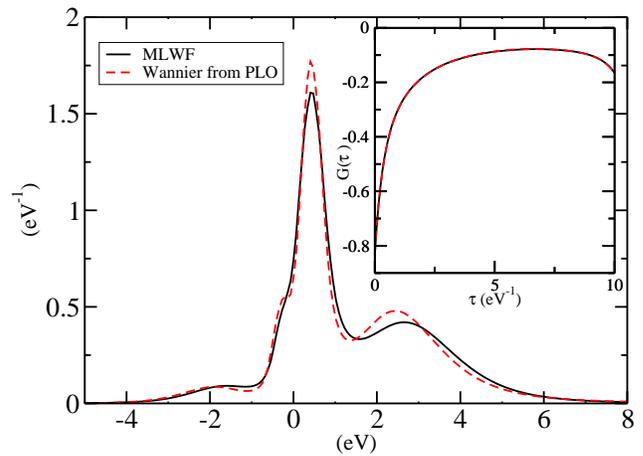}}}}
\caption{(Color online) Local impurity
spectral function of SrVO$_3$ for U=4 eV within LDA+DMFT using the MLWF basis and
the PLO scheme ($N_{b}=3$). Inset: Green's function in imaginary time. Note that when
using such a small ($t_{2g}$) energy window, the impurity spectral function and Bloch-resolved
spectral function coincide.
}
\label{fig:nw3}
\end{figure}

Spectral functions obtained from the PLO and MLWF schemes are reproduced on figure \ref{fig:nw3}.

The impurity Green's function obtained within the two schemes are in very good agreement: they both
show a lower Hubbard band at -1.8 eV and a upper Hubbard band at 2.5 eV.
This shows that the two schemes contain the same physical content and that PLOs constructed from
a small energy window give very similar results to
{\it maximally localized} Wannier functions~\cite{Lechermann06}.

\subsubsection{Oxygen-$p$ and Vanadium-$t_{2g}$ bands in the basis for $\mathcal{W}$ ($N_{b}=12$).}
Oxygen $p$ states are now included in the calculation.  MLWFs of $t_{2g}$ symmetry will
thus be more localized, and will have less O-$p$ character, because the MLWFs of Oxygen $p$ symmetry
will have mainly the weight on Oxygens atoms. In the PLO scheme also, the Wannier orbitals
will include more Bloch states, and thus will be closer to
to localized local orbitals $\chi_m$.
In other words, because the Wannier function of the local impurity problem is now constructed from
a wider energy windows which includes oxygen $p$ states, the corresponding spectral function will
have a non negligible weight in the energy area of the oxygen-$p$ states. In the LDA case without Hubbard
correction, this spectral function would in fact be identical to the projected $t_{2g}$ density of states
plotted on Fig \ref{fig:bsSrVO3}.

\begin{figure}
{\resizebox{8.0cm}{!}
{\rotatebox{0}{\includegraphics{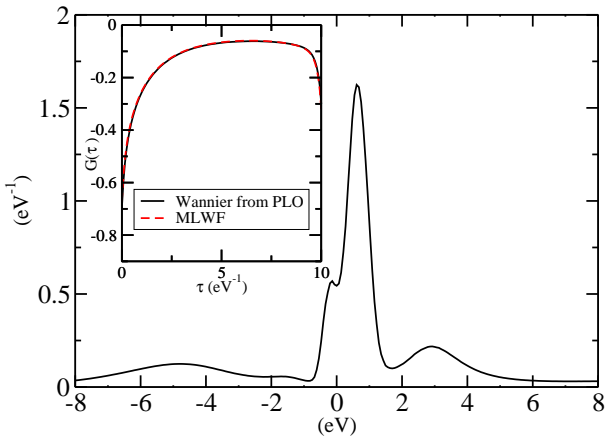}}}}
\caption{(Color online) Local impurity spectral function of SrVO$_3$ for U=6 eV within LDA+DMFT using
the PLO scheme ($N_{b}=12$). Inset: Green's function in imaginary time for both the PLO and the MLWF schemes.}
\label{fig:nw12bench}
\end{figure}

In order first to benchmark our calculation using PLO with respect to calculations using MLWF,
we plot on figure \ref{fig:nw12bench} the comparison of the impurity Green's function
in the two cases.
The local impurity spectral function shows a large hybridization band
in the area of the oxygen $p$ states. We emphasize that
this band is mainly not a many-body feature: it is a manifestation of hybridization with
oxygen states, and is readily visible at the pure LDA level.
The lower Hubbard band is in fact partly contained
in this spectral function as a shoulder in the hybridization band around -1.5 eV.
However, disentangling the Hubbard band from the hybridization contribution is somewhat difficult
to achieve with the Maximum entropy method. The upper Hubbard band is visible
in the impurity spectral function, as a hump around 3.0 eV.

\begin{figure}
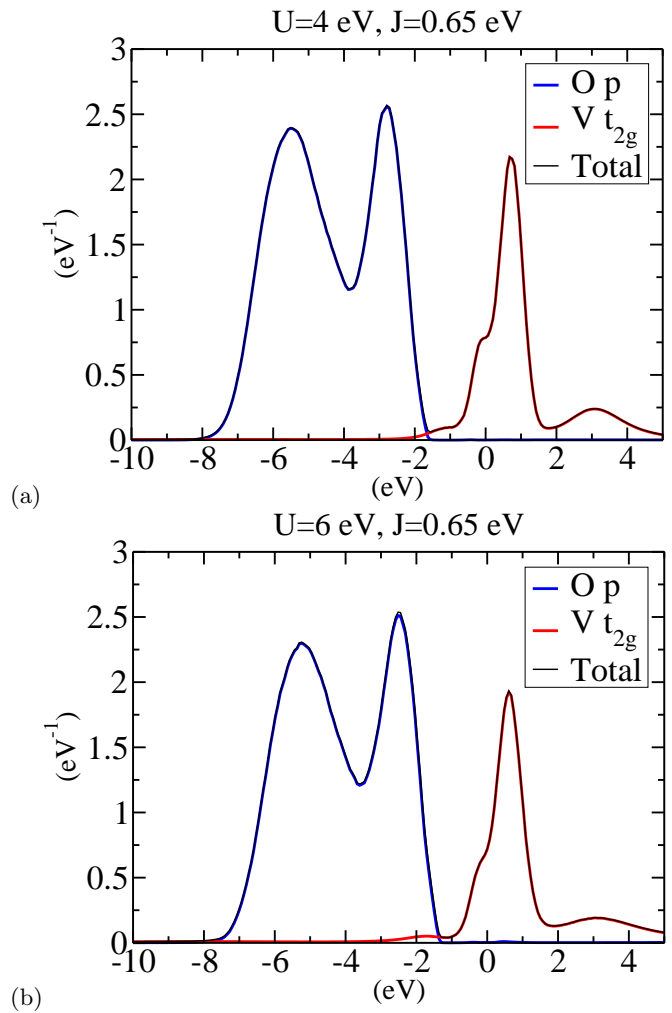

(a){\resizebox{8.3cm}{!}
{\rotatebox{0}{\includegraphics{dosbands.U=4.t2gOxy.eps}}}}\\
(b){\resizebox{8.3cm}{!}
{\rotatebox{0}{\includegraphics{dosbands.U=6.t2gOxy.eps}}}}
\caption{(Color online) Total Bloch spectral function for SrVO$_3$, and spectral functions
of Oxygen-$p$,  and Vanadium-t$_{2g}$ \emph{states}
for U=4 eV (a) and U=6 eV (b) within LDA+DMFT using the
PLO scheme ($N_{b}=12$).
These spectral functions are not
the local orbitals projected spectral functions (see text): the $t_{2g}$ local impurity spectral functions
are plotted on figure \ref{fig:nw12bench}.}
\label{fig:nw12plo}
\end{figure}

These many-body features (lower and upper Hubbard band) are more clearly revealed,
however, by looking at the ($\bk$-averaged) spectral function of $t_{2g}$
Bloch states (\ref{eq:Abloch}). It is also useful to look at the spectral functions of the
other Bloch states in the basis-set. The summation of these
spectral functions over bands belonging to the same group (e.g Oxygen-$p$ or Vanadium-t$_{2g}$ states,
ie bands with mostly  Oxygen-$p$ or Vanadium-t$_{2g}$ character )
enables us to have a clear view on the impact of correlation on LDA bands.
The spectral function of Oxygen-$p$ {\emph states},  and Vanadium-t$_{2g}$ {\emph states}
are plotted on figure  \ref{fig:nw12plo}.
We emphasize that these spectral functions do not correspond to atomic-like orbitals with
Oxygen-$p$ character and Vanadium-t$_{2g}$ character (the latter being the impurity spectral
function plotted on Figure \ref{fig:nw12bench}).

The Hubbard band appears as a hump in the $\bk$-averaged Bloch spectral function
corresponding to the $t_{2g}$ states for U=4 eV.
This hump is located between the O-$p$ states and the $t_{2g}$-states quasiparticle peak.
The Hubbard band is more clearly resolved for U=6 eV. In this case, however, it is
hidden inside the Oxygen $p$ band. We shall come back to this point at the end of this section.

The fact that a higher value of U is necessary in this case (with respect to $N_{b}=3$.)
to recover the lower Hubbard band is consistent with the fact that Wannier functions
are more localized for $N_{b}=12$.

\subsubsection{Large number of bands in the basis for $\mathcal{W}$ ($N_{b}=21$)}

\begin{figure}
{\resizebox{8.3cm}{!}
{\rotatebox{0}{\includegraphics{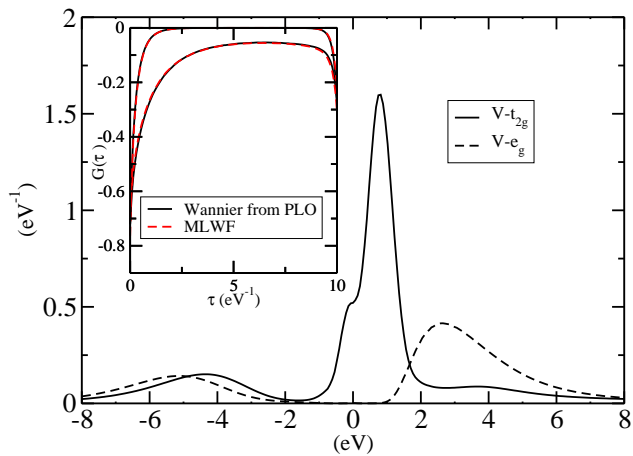}}}}
\caption{(Color online) Local impurity spectral function of SrVO$_3$ for U=6 eV within LDA+DMFT using
the PLO scheme ($N_{b}=21$). Inset: Green's function in imaginary time for both the PLO and the MLWF schemes.}
\label{fig:nw21bench}
\end{figure}

In this case, the $e_g$ states are included in the calculation. The
impurity model is now solved with all five $d-$ orbitals.
The agreement between impurity Green's functions computed in the MLWF
and PLO schemes is shown on figure \ref{fig:nw21bench}. Note again,
that resolving the Hubbard band from the impurity Green's function is difficult because
this quantity is dominated by hybridization effects with oxygen states.
Again, we have to turn to Bloch-resolved spectral functions,
plotted in Fig.~\ref{fig:nw12plo} for
Oxygen-$p$,  Vanadium-t$_{2g}$  and Vanadium-e$_g$ Bloch states.
The results are quite similar to the previous ones with $N_{b}=12$, with a
lower Hubbard band clearly visible for the $t_{2g}$ states
at about -2.0 eV in the spectral function. Note that here we use $U=6$ eV.
It shows that the basis of Kohn-Sham bands
is adapted to the calculation: the convergence of physical properties
as a function of the number of bands is rather fast.
\begin{figure}
{\resizebox{8.3cm}{!}
{\rotatebox{0}{\includegraphics{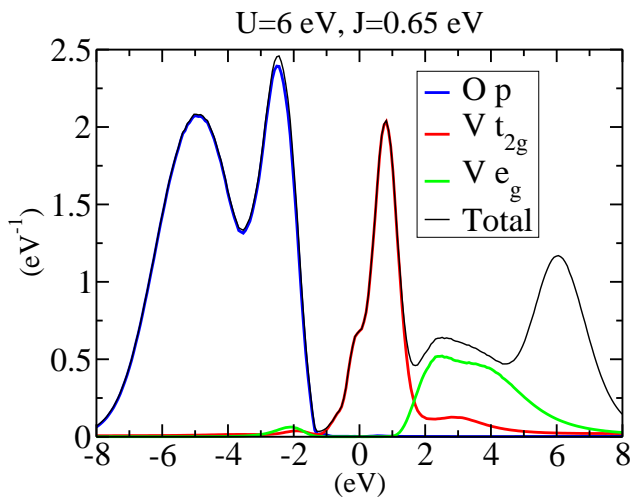}}}}
\caption{(Color online) Total Bloch spectral function for SrVO$_3$, and spectral functions
of Oxygen-$p$,  Vanadium-t$_{2g}$ and Vanadium-e$_g$ {\emph states}
for U=6 eV within LDA+DMFT using the
PLO scheme ($N_{b}=21$).
These spectral functions are not
the local orbitals projected spectral functions (see text): the $t_{2g}$ and $e_g$ local impurity spectral functions
are plotted on figure \ref{fig:nw21bench}.}
\label{fig:nw21plo}
\end{figure}

\subsubsection{Discussion}

The comparisons made above should not be simply thought of as
a convergence study as a function of the size of the Bloch basis ${\cal W}$. Indeed,
as $N_b$ is increased, we change the spatial extension of the local orbitals (Eq. \ref{eq:wkfm}) spanning the
correlated subspace, so that the DMFT treatment does not apply to the same objects.
Convergence studies with fixed $\chi_m$'s could also be performed, but this is not the
main scope of this article.

\begin{figure}
{\resizebox{7.0cm}{!}
{\rotatebox{0}{\includegraphics{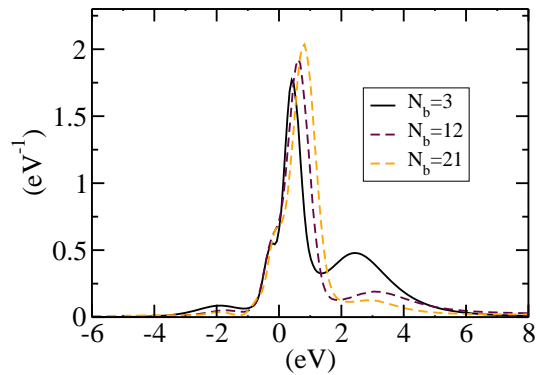}}}}
\caption{(Color online) Spectral function of $t_{2g}$ Bloch states for SrVO$_3$ within the PLO scheme of LDA+DMFT.
For $N_{b}=12$ and
$N_{b}=21$: U=6eV. For
$N_{b}=3$: U=4eV.}
\label{fig:nwcompa}
\end{figure}

Instead, we would like to emphasize some of the physical issues when applying DMFT to
different local orbitals with different degrees of spatial localization.
As we have seen, for more localized orbitals (larger $N_b$), we have to increase the value
of the on-site $U$ on the vanadium site to get consistent results. This is of course expected
physically. We see however that the main features of the spectral function
(see Fig. \ref{fig:nwcompa}) are almost identical for $N_b$=12 and
$N_b$=21 with similar values of $U$.  The lower Hubbard band (between -1.7 eV and -2.0 eV), is not far from
the position found experimentaly\cite{sek04,yos05,mai05,wad06,egu06} ( between  -1.5 eV and -2.0 eV ).
The Hubbard band (between 2.5 and 3eV) is also in the range of experiments (2.5 eV).

\begin{figure}
{\resizebox{7.0cm}{!}
{\rotatebox{0}{\includegraphics{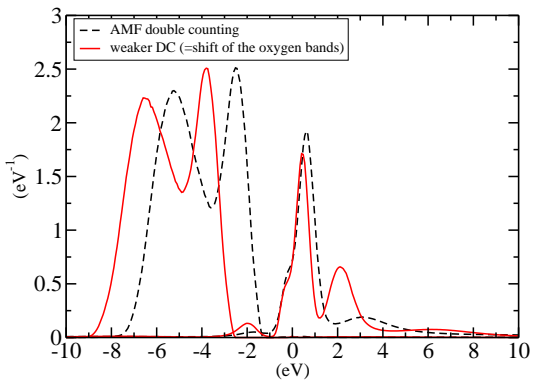}}}}
\caption{(Color online) Spectral functions
of Oxygen-$p$,  and Vanadium-t$_{2g}$ \emph{states}
for U=6 eV within LDA+DMFT using the
PLO scheme ($N_{b}=12$). Two values of
the double counting shift are used:  6.6 eV (AMF) and 3.6 eV.}
\label{fig:dcweak}
\end{figure}

In the results that we have obtained, the lower Hubbard band is rather systematically at an energy
in which oxygen states already give a sizeable contribution in the total density of states.
Experiments, however, seem to suggest a somewhat larger separation. We believe that this is due
to the fact that the relative location of oxygen and V-$t_{2g}$ state is not accurately obtained at the
LDA level, and that a better starting point (such as GW) is required to handle this problem with
better accuracy. Indeed, we have verified that the intensity of the lower Hubbard band, and
especially, of the upper Hubbard band (as revealed in the Bloch-resolved $t_{2g}$ spectral function)
is very sensitive to the precise value of the double-counting correction which is used.
This is shown in Fig.~\ref{fig:dcweak}, in which we have slightly shifted downwards the O$-2p$
states with respect to the V-$t_{2g}$ ones by choosing purposedly a smaller value of the double counting correction
(3.6 eV instead of 6.6 eV).
The more isolated the $t_{2g}$ states are, and the more prominent are correlation effects within that band (for
a given $U$).

In future work, calculations with extended basis sets should naturally face the issue of
calculating the on-site Coulomb interaction from first-principles, but also of taking into account
Coulomb repulsion terms $U_{pp}$ on the oxygen sites, as well as inter-site repulsions $U_{pd}$ between
vanadium and oxygen. We note that treating the latter in the Hartree approximation precisely brings
in a correction to the relative position of oxygen and vanadium states, of the same nature
than the double-counting terms.

The general conclusion of this study of SrVO$_3$ at this stage is that our formalism is able to describe the main
features of the experimental function within a general formalism which can
take into account all states in the basis.

\section{Application: $\beta$-${\rm Ni}{\rm S}$}
\label{sec:nis}

\begin{figure}[t]
\includegraphics*[height=5cm]{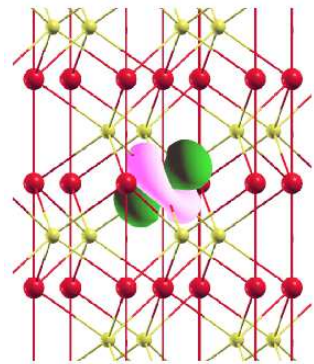}
\caption{(Color online) Projected $\beta$-NiS structure with a local Ni($e_g$)
orbital obtained from a MLWF construction using the block of 16 Ni$(3d)$/S$(3p)$
bands. The $c$ axis is identical to the vertical axis.
\label{nisstruc}}
\end{figure}

The hexagonal form of nickel sulfide ($\beta$-NiS) has attracted a lot of interest
over the years since it exhibits a first-order electronic phase
transition~\cite{Spa67} at about 260 K. Whereas the high-temperature phase may be
classified as a paramagnetic metal, below the transition $\beta$-NiS shows
antiferromagnetic order and the resistivity behavior corresponds to characteristics
of a semi metal~\cite{Whi71} or a degenerate semiconductor~\cite{Oht70}. Hence the
term ``antiferromagnetic nonmetal'' is commonly used for the low-temperature phase.
Note that hexagonal NiS is only metastable at room temperature, the true stable
phase is given by the millerite structure~\cite{Ben55}. The crystal structure
(Fig.~\ref{nisstruc}) of $\beta$-NiS is of the NiAs-type (space group $P6_3/mmc$)
with two unit cells in the primitive cell~\cite{Tra70}. In this rather simple
structure the NiS$_6$ octahedra share edges within the $ab$ plane and share faces
along the $c$ axis. There is a slight decrease of the cell parameters
($\delta a/a$$\sim$0.3\%,$\delta c/c$$\sim$1\%) below the transition, giving rise
to a volume collapse of the order of 2 percent. Our aim in this work is not to
perform a detailed study of the metal-to-nonmetal transition of $\beta$-NiS,
since this would involve deeper considerations concerning the role of magnetism
above and below the transition. For an overview of this, still controversially
discussed, topic see e.g. the review by
Imada {\sl et al.}~\cite{imada_mit_review}.

Here we mainly want to use the high-temperature phase of $\beta$-NiS as an example
for a correlated metal with strongly hybridized Ni($3d$) and S($3p$) states, formally
in the charge-transfer regime of the Zaanen-Sawatzky-Allen classification
scheme~\cite{zsa_1985_prl}. In this regard it will become
clear that this compound may not easily be treated in the ``traditional'' LDA+DMFT
scheme via projecting solely onto low-energy states close to the Fermi level. Thus
the goal is to use the here outlined projection technique of interfacing LDA with
DMFT in order to explore the importance of electronic correlations for the local
spectral function.

\subsection{LDA investigation}

\begin{figure}[b]
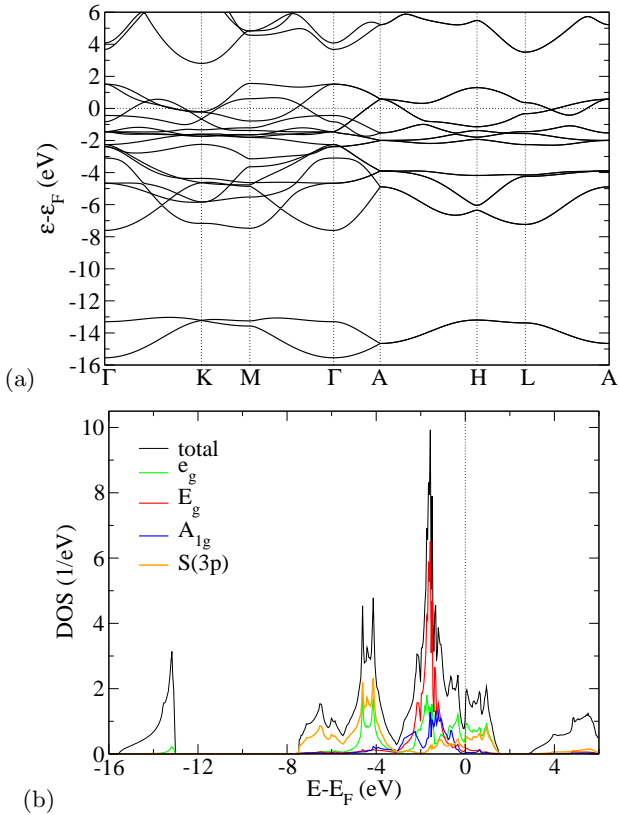

(a)\includegraphics*[width=7.75cm]{nis.ldabands.eps}\\[0.2cm]
(b)\includegraphics*[width=7.25cm]{nis.ldados.eps}
\caption{(Color online) LDA data for $\beta$-NiS.
(a) Band structure. (b) DOS. For the local Ni$(3d)$/S$(3p)$-DOS the cutoff radius
was half the nearest-neighbor distance, respectively.\label{nislda}}
\end{figure}
\begin{figure}[t]
\includegraphics*[width=7.5cm]{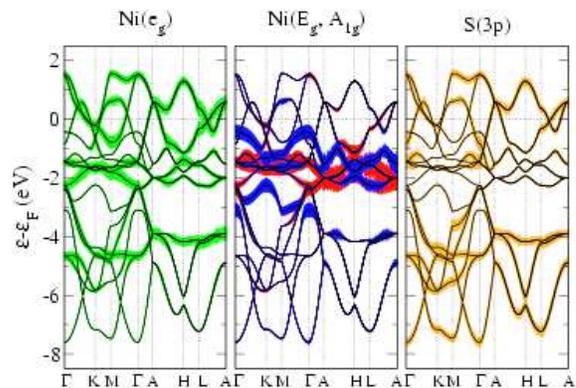}
\caption{(Color online) LDA fatband decomposition for $\beta$-NiS. (red/gray)
Ni($E_g$) and (blue/dark) Ni($A_{1g}$). The local Ni$(3d)$/S$(3p)$ projection
cutoff radius was half the nearest-neighbor distance, respectively.\label{nisldafat}}
\end{figure}

Electronic structure calculations for $\beta$-NiS date back to the original
work of Mattheis~\cite{mat74}. Here the DFT-LDA calculations are performed
with a mixed-basis pseudopotential (MBPP)
code~\cite{mbpp_code} described in appendix~\ref{app:mbpp}. For the lattice
parameters we use $a$=3.440 \AA~ and $c$=5.351 \AA~ from Ref.~\onlinecite{Tra70}.
Figure~\ref{nislda} shows the resulting band structure and density of states (DOS)
from this computation. It is seen that the common block of hybridized Ni($3d$) and
S($3p$) bands is isolated from the remaining bands. The dominantly S($3s$)-like bands
are $\sim$5.5 eV below in energy~\cite{huf85}, while the bands above starting with
mainly Ni($4s$) character are separated by $\sim$1.2 eV from that block. The latter
has a width of $\sim$9 eV and consists of three prominent peaks below the Fermi
level. The dominant $d^8{\bf L}$ peak~\cite{huf85} at -1.5 eV (P$_1$), a second at
-4.4 eV (P$_2$) and the last at -6.5 eV (P$_3$). Whereas P$_2$ has mixed
Ni($3d$)/S($3p$) character, P$_3$ is a nearly pure S($3p$) peak and does not
stem from bonding between Ni and S.

In order to obtain more detailed information about the involvement of the different
orbital sectors, the orbital-resolved DOS is additionally incorporated in
Fig.~\ref{nislda}b. Furthermore, the fatband representation of the decomposition of
the Bloch bands is presented in Fig.~\ref{nisldafat}. Because of the hexagonal
symmetry the Ni($3d$) multiplet splits into two degenerate $e_g$ levels,
two degenerate $E_g$ levels, and one $A_{1g}$ level per atom, respectively. As usual,
the $e_g$ states hybridize more strongly with the S($3p$) states than the remaining
($E_g$,$A_{1g}$) states. Due to the increasingly filled $d$ shell of Ni, the
($E_g$,$A_{1g}$) levels are moreover also nearly completely occupied in $\beta$-NiS.
In numbers, the LDA filling, including degeneracy, amounts to (2.9;3.9,1.9) for
($e_g$;$E_g$,$A_{1g}$).
The fatband representation clearly shows that the hybridization between the different
orbitals are strong and its not quite obvious to single out a low-energy regime for
this compound within the central block of bands. Although the bands at the Fermi
level appear to be dominantly of $e_g$ character, the correspondig orbitals have
strong weight at lower energy, too. Despite the strong filling of the ($E_g$,$A_{1g}$)
states, abandoning those orbitals in a minimal hamiltonian description of $\beta$-NiS
is likely to fail. Fluctuations originating from hole-like states may be important in
the end to reach a better understanding of the complex magnetic behavior of this
material. Note also in this respect that the effective bandwidth of the energy range
where the $e_g$ character is manifest is roughly twice as large as the corresponding
energy range for ($E_g$,$A_{1g}$). Hence this reduced relative bandwidth for
($E_g$,$A_{1g}$) within the Ni($3d$) multiplet will also have influence on the
degree of the orbital-resolved correlation effects.

\subsection{LDA+DMFT investigation}

The influence of correlation effects in $\beta$-NiS has been investigated,
experimentally and theoretically, by several
authors~\cite{huf85,fuj88,Anisimov91,nak94b,usu00,Kri00}.
It is generally believed that this compound is moderately correlated in view of the
strong hybridization of the $\sim$$3/4$ filled $e_g$ states with sulfur. A value
for the Hubbard $U$ of the order of 4$-$5 eV was estimated from modelling the
measured spectral function derived from photoemission
experiments~\cite{fuj88,nak94b,Kri00}. These experiments in the metallic phase are
in rough agreement with the shown LDA DOS below the Fermi level in so far as they
also reveal three peaks with comparable relative intensity as the theoretical set
(P$_1$, P$_2$, P$_3$). However, an effective
single-particle picture appears to be insufficient to understand those peaks,
especially when going to the nonmetallic phase~\cite{nak94b}. It is theoretically
expected that a lower Hubbard band, i.e., satellite, originating from the Ni($3d$)
states should be located within the Bloch states deep in energy (starting around 6
eV) and dominantly characterized by S($3p$). This idea relies on the fact that
$\beta$-NiS may be viewed as belonging to the charge-transfer category of
transition-metal chalcogenides, yet not being that strongly correlated for the lower
Hubbard band appearing completely below the S($3p$) states.

\begin{figure}[b]
\includegraphics*[width=7.5cm]{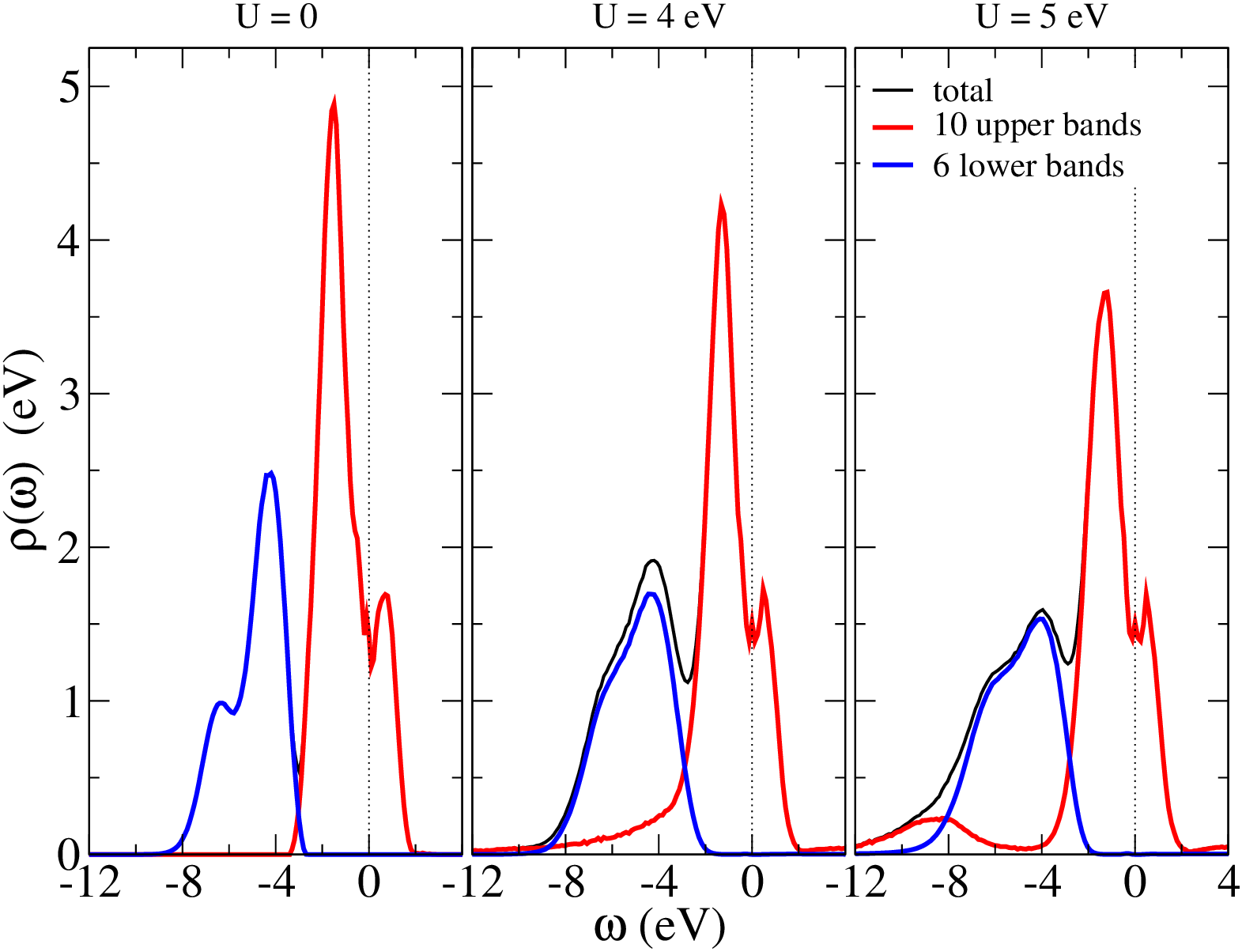}
\caption{(Color online) The LDA+DMFT local spectral function for $\beta$-NiS derived
from the local Green's function in Bloch basis at $\beta$=10 eV$^{-1}$.
The contribution from the upper 10 bands of the Ni($3d$)/S($3p$) block was encoded
red, the one from the lower 6 bands was encoded blue. This guidance to the eye should
roughly seperate dominant Ni($3d$) from dominant S($3p$) character.\label{nisldadmft1}}
\includegraphics*[width=7.5cm]{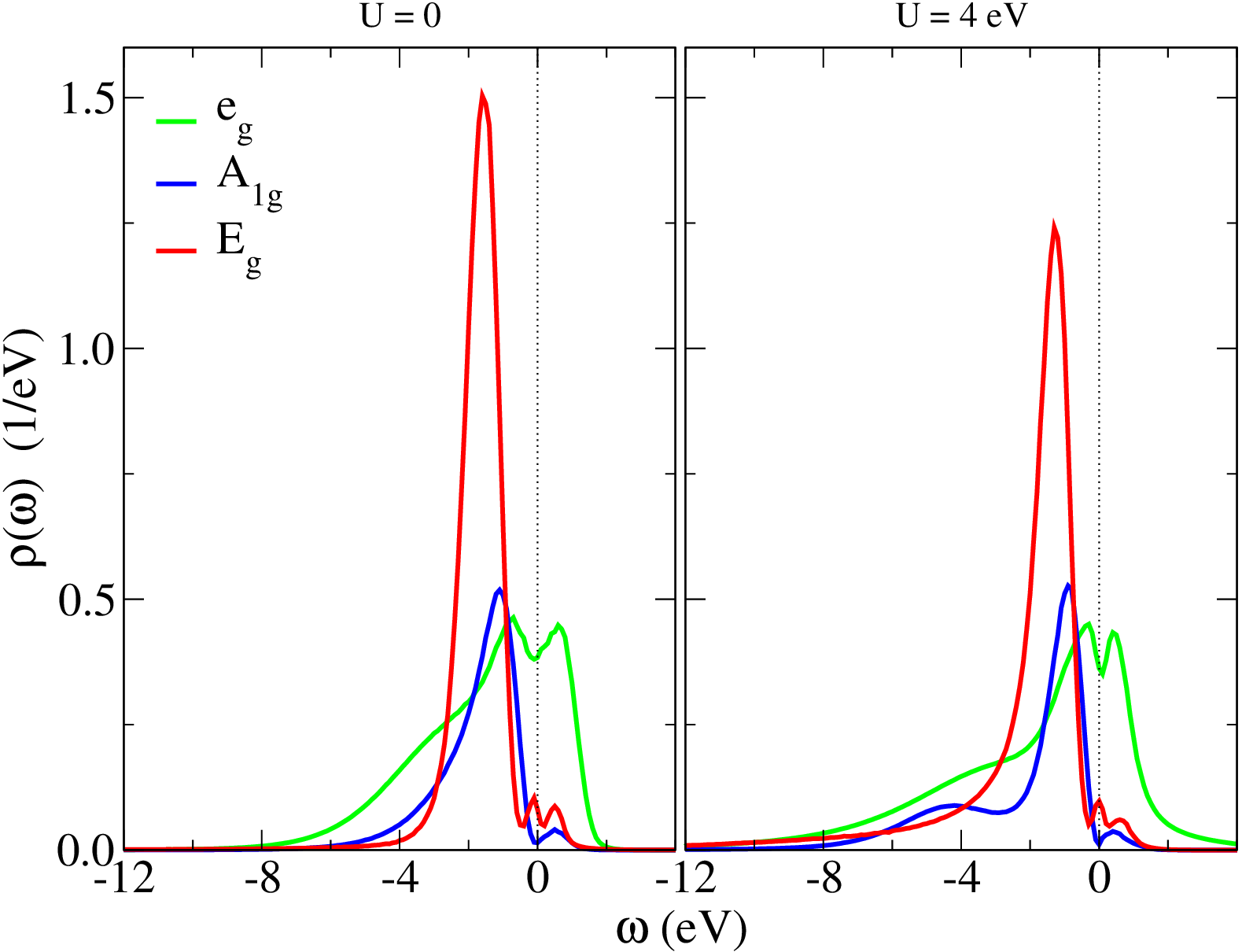}
\caption{(Color online) The LDA+DMFT local impurity spectral function for
$\beta$-NiS at $\beta$=15 eV$^{-1}$.\label{nisldadmft2}}
\end{figure}

The strong hybridization between Ni($3d$) and S($3p$) renders the PLO version of
the LDA+DMFT method most suitable for this compound. Concerning the value of
$U$ we chose a pragmatic approach and performed the calculations for two possibly
reasonable choices, i.e., $U$=4 eV and $U$=5 eV. The value of $J$ is certainly less
materials dependent and was fixed to $J$=0.7 eV. In order to take care of the double
counting, we used the formalism of fixing the local total charge (see
appendix~\ref{dc-app}). Note that for the present crystal structure there are two
symmetry-equivalent Ni atoms in the primitive unit cell, i.e., two correlated sites
$\bR$. The local Green's function and self-energy were thus computed by symmetrizing
the site-resolved quantities. For the projection onto local orbitals we limited
the number of bands to $N_b$=16, i.e., all bands of the central block around the
Fermi level are used. This renders the corresponding correlated subspace already
rather localized, e.g., the effective Ni($3d$)-like WFs from this set of bands are
not expected to leak much to the sulfur sites (see also Fig.~\ref{nisstruc}).

By explicitly including the correlation effects, the orbital-resolved fillings
in the $3d$ shell of Ni do not change relevantly, thus effects due to changes in
orbital populations induced by correlations are
not expected to play an important role for this compound. Figure~\ref{nisldadmft1}
exhibits the resulting local spectral function for the different values of $U$ at
inverse temperature $\beta$=10 eV$^{-1}$. It is seen that the influence of
correlation effects on the metallic spectral function are indeed rather subtle for
this compound. A lower Hubbard band appears to show up for $U$$>$4 eV within the
dominant S($3p$) energy regime. It is however hard to extract this atomic-like
excitation from the total spectral function (as in experiment~\cite{huf85,Kri00}).
The plot of the {\sl impurity} spectral function in Fig.~\ref{nisldadmft2} reveals
that the ($E_g$,$A_{1g}$) orbitals yield indeed effective bands with smaller
bandwidth than the $e_g$ orbitals. It seems that the contribution to the lower
Hubbard band then also stems more significantly from the former set of orbitals.

In the end, the total spectral function computed for $U$=5 eV shows close resemblance
to recent experimental curves obtained from photoemission~\cite{Kri00}. Note however
that it appears tricky to disentangle the lower Hubbard band from the $P_3$ peak of
in principle pure S$(3p)$ content. Our calculation suggests that the correlation
effects have also impact on the lowest S$(3p)$-like states. Further
studies are necessary to clarify this issue.

\section{Conclusions}

We have implemented an effective and flexible LDA+DMFT scheme into two electronic
structure formalisms based on a plane-wave description, i.e., projector 
augmented-wave and mixed-basis pseudopotential.
The orbitals defining the correlated subspace ${\cal C}$,
in which many-body effects are included, are constructed by projecting local atomic-like
orbitals onto a restricted set ${\cal W}$ of Kohn-Sham Bloch states, similar to the procedure adopted in
Ref.~\onlinecite{Anisimov05}. Orthonormalization of the projected orbitals
yields effective Wannier functions spanning ${\cal C}$.
These WFs are not unique: they depend  on the energy window covered by the Bloch functions
in ${\cal W}$ (the larger the window, the more localized the WFs are).
Although more sophisticated constructions~\cite{Marzari97,Andersen_NMTO_00}
of explicit WFs will surely remain an important tool in the LDA+DMFT context, we
feel that the more straightforward projection technique will render future
developments in this area easier and more efficient.

Using this method, we have investigated SrVO$_3$ and compared different implementations
involving different choices for the set of Bloch states ${\cal W}$ and for the corresponding
local orbitals spanning ${\cal C}$. We have shown that the basic physical findings for
this compound are consistent with previous LDA+DMFT treatments, independently of the
chosen implementation. However, the present study
allows for an explicit treatment of ligand states, which raises several issues which
should be the subject of further studies. One of these issues is a first-principle
determination of the on-site Coulomb matrix elements in the different choices of local
orbitals and basis-set. Our study supports the expected fact that a larger value of
$U_{dd}$ has to be taken when a larger energy range (and more localized orbitals) is
considered. Furthermore, a proper treatment of the on-site repulsion on oxygen sites $U_{pp}$,
as well as of oxygen-transition metal inter-site repulsion $U_{pd}$ should certainly be
considered. This issue is tightly connected to the choice of double-counting correction, which, as
we have shown, plays a significant role.

We have also presented a first LDA+DMFT study of metallic $\beta-$NiS, a charge-transfer compound for
which the inclusion of ligand states is crucial.  More detailed
studies, extending also into the nonmetallic regime, are needed in order to gain more insight
into this ``traditional companion'' of the more famous NiO compound
where such investigations recently took place~\cite{kun07a,kun07b}.

In general, the explicit inclusion of the ligand states in the LDA+DMFT description,
based on generic highly-accurate electronic structure codes, will open the door
to new possibilities for the investigation of strong-correlation effects in real
materials.

\appendix
\section{Calculation of the projection $P^{\bR}_{m\nu}$}
\label{appendix:psichi}
In this appendix, we describe the details of the calculation
of the projection $P^{\bR}_{m\nu}$ from KS Bloch states onto local orbitals in both
the PAW\cite{Blochl94} and the MBPP~\cite{lou79} method.

\subsection{Projected Augmented-Wave (PAW) method}

In the projected augmented-wave formalism, two kinds of atomic functions are used:
the pseudowavefunction $\tphi$ and the true wavefunction $\varphi$. They are used to
recover the correct nodal structure of the wavefunction near the nucleus
(see Eq. (9) of Bl\"ochl {\it et al}\cite{Blochl94}). A direct calculation
of the projection $P^{\bR}_{m\nu}(\bk)$  would thus require to compute the three
terms resulting from this equation, namely ($\tp_i$ is the projector $n_i$ for
angular momenta $l_i$, and its projection $m_i$):
\begin{equation}
P^{\bR}_{m\nu}(\bk)=
\langle \chi_m^{\bR}| \tPsi_{\bk\nu} \rangle
+ \sum_i \bra \tp_i | \tPsi_{\bk\nu} \ket
( \bra \chi_m | \varphi_i \ket -\bra \chi_m^{\bR} | \tphi_i \ket )
\label{eq:pawproj}
\end{equation}
However, atomic $d$ or $f$ wavefunctions are mainly localized inside spheres.
As a result,
we can compute the projection only inside sphere, in the spirit of current LDA+U
implementation~\cite{Bengone00,Amadon07b} in PAW.
In this particular case, the first and third terms of Equation \ref{eq:pawproj},
cancel with each others. This cancellation is only exact for
a complete set of projectors in the energy range of the calculation.
This completeness can be easily tested during the construction of projectors
and partial waves.
The projection thus writes as a sum over projectors $\tp_{n_i}$:
\begin{equation}
P^{\bR}_{m\nu}(\bk) = \sum_{n_i} \bra
\tp_{n_i} | \tPsi_{\bk\nu} \ket
\bra \chi_m^{\bR} | \varphi_{n_i} \ket
\end{equation}
We note that integrals in this equation are computed only
inside sphere. It implies that the projection is done on an unnormalized wavefunction
$\chi$. Nevertheless, the later normalization of the projection $P^{\bR}_{m\nu}(\bk)$
will make a normalization of $\chi$ redundant. This implementation has been made
with the code ABINIT\cite{abinit1,torrent07}.

Additionally, we implemented the
proposed LDA+DMFT scheme in a somehow simplified interfacing with the PAW-based
Vienna Ab Initio Simulation Package~\cite{kre94} (VASP) by using {\sl only} the PAW
projectors from the standard output for $P^{\bR}_{m\nu}(\bk)$ . Although approximate,
this latter identification yielded (after, of course, proper normalization) very
similar results in comparison with the more well-defined explicit coding described
above.

\subsection{Mixed-Basis pseudopotential (MBPP) method\label{app:mbpp}}

The combination of normconserving pseudopotentials with a mixed basis of plane waves
and localized orbitals in order to represent the pseudo crystal wave function is the
main ingredient~\cite{lou79} of our employed mixed-basis pseudopotential
code~\cite{mbpp_code}.
More concretely, the MBPP band structure code uses the following representation for
the KS pseudo wave functions:
\begin{equation}
\Psiknu=\sum_{\bG}\psiG\ketkG+
\sum_{\g lm}\beta^{\g l}_{\bk\nu m}\phik\qquad,\label{mbpp_basis}
\end{equation}
with
\begin{eqnarray}
\langle\br\ketkG&=&\frac{1}{\sqrt{\Omega_c}}
{\rm e}^{i(\bk+\bG)\cdot\br}\\
\langle\br\phik&=&\sum_{\bT}{\rm e}^{i\bk\cdot(\bT+\bR_\g)}
\phi^{\g l}_m (\br-\bT-\bR_\g)\,\,.
\end{eqnarray}
In these eqns., $\g$ denotes the atom in the unit cell, $\Omega_c$ the volume of
the unit cell, $\nu$ the band and $l$,$m$ are angular momentum and azimutal
quantum number. The plane waves in this basis extend up to a chosen energy cutoff
$E^{\rm pw}$.
The analytical form of the local orbitals $\phi^{\g l}_m$ reads
\begin{equation}
\phi^{\g l}_m(\br')=i^l f_{\g l}\,K_{lm}(\hat{\br}')\,\,,\quad
\br'=\br-\bT-\bR_\g\quad,\label{mbpp_locorb}
\end{equation}
whereby $K_{lm}$ descibes a cubic harmonic and the radial function $f_{\g l}$ is an
atomic pseudo wave function modified with a proper decay and cutoff function in
order for $f_{\g l}$ to vanish beyond some chose radial cutoff $r_{c}^{(\g l)}$.
Note that the local orbitals are orthonormal and by definition do not
overlap between neighboring sites, i.e.,
\begin{eqnarray}
\int_0^{r_{c}^{(\g l)}}dr |f_{\g l}(r)|^2\,r^2&=&1\\
\langle\phi^{\g' l'}_{m'}|\phi^{\g l}_{m}\rangle&=&
\delta_{\g'\g}\delta_{l'l}\delta_{m'm}
\,\,.\label{mbpp_locortho}
\end{eqnarray}
By identifying the local orbitals $\phi^{\g l}_m$ as the objects to project onto
(i.e. the $\chi_m$ in the outlined LDA+DMFT formalism), the
projection $P^{\bR}_{m\nu}(\bk)$ may be written with the help of
(\ref{mbpp_basis}) and (\ref{mbpp_locortho}) as
\begin{equation}
P^{\bR_\g}_{m\nu}(\bk)=\langle\phi^{\g l}_{\bk m}\Psiknu=
\sum_{\bG}\psiG\langle\phi^{\g l}_{\bk m}\ketkG+\beta^{\g l}_{\bk\nu m}\quad.
\end{equation}

\section{Double counting\label{dc-app}}
The double-counting shift (used in eq. \ref{eq:scc_self}) is necessary in the
formalism in order to correct the fact that electronic correlations are already
treated in an average way within LDA. Since there is no unique way to extract this
double counting from LDA, two alternative formulations were used in this work.
First the so-called Around Mean Field~\cite{Anisimov91,Czyzyk94} (AMF) method, which
is usually adequate for metals. Second, a different formalism in which the
double counting is such that the electronic charge computed from the local
noninteracting Green's function and the one computed from the interacting impurity
Green's function is constraint to be identical:
\begin{equation}
{\rm Tr}\, G_{\rm imp}(\iomn)={\rm Tr}\, G_{\rm imp}^{(0)}(\iomn)\qquad.\label{gfix}
\end{equation}
Here $G_{\rm imp}^{(0)}$ is generally given by
\begin{eqnarray}
G^{\bR,\rm{imp},(0)}_{mm'}(\iomn) &=&
\sum_{\bk}\sum_{\alpha\alpha'}
\langle\chi^{\bR}_{\bk m}|\Bka\rangle\langle\Bkap|
\chi^{\bR}_{\bk m'}\rangle
\times \nonumber \\
&&\hspace{-2cm}\times\left\{\left[\iomn+\mu-\bH_{\rm{KS}}(\bk)
\right]^{-1}\right\}_{\alpha\alpha'}\,.
\end{eqnarray}
Alternatively one may want to use the Weiss field ${\cal G}(\iomn)$ instead of
the local noninteracting Green's function in eq. (\ref{gfix}). Note that in any case
this formalism asks for an additional convergence parameter $\delta\mu_{dc}$ to
fulfill (\ref{gfix}). Hence this parameter has to be found iteratively together with
the total chemical potential $\mu$ in the DMFT cycle.

For SrVO3 (with $N_b=12$), the respective double-counting shifts are 6.6 eV (AMF) and
5.5 eV (fixing local charge). However these two calculations give very similar
results.

\section{Maximally Localized Wannier Functions in PAW}
\label{app:Wannierpaw}
To define maximally-localized~\cite{Marzari97} WFs, we need the following quantity:
\begin{equation}
M_{m,n}^{(\vk,\vb)}=\bra \Psi_{n,\vk}|{\rm e}^{-i\vb.\vr}|\Psi_{m,{\bf k+b}} \ket
\end{equation}

In the PAW framework, this quantity can be expressed as a function of the
pseudo wave function $\widetilde{\Psi}$, the projectors $p$, the atomic wave
function $\phi$ and the pseudo atomic wave function $\widetilde{\phi}$.
We use the expression for an operator $A$ within the
PAW formalism (Eq. (11) of  Bl\"ochl {\it et al}\cite{Blochl94}).
We thus obtain:
\begin{eqnarray}
\nonumber
M_{m,n}^{(\vk,\vb)}&=&\bra
\widetilde{\Psi}_{n,\vk}|{\rm e}^{-i\vb.\vr}|\widetilde{\Psi}_{m,{\bf k+b}} \ket \\
&+& \sum_{i,j}  \bra \widetilde{\Psi}_{\vk,m}|\widetilde{p}_i \ket
\bra \widetilde{p}_j|\widetilde{\Psi}_{\vk+\vb,m} \ket \\
\nonumber
&\times&
\left(
\bra \phi_i |{\rm e}^{-i\vb.\vr}|\phi_j \ket -\bra \widetilde{\phi}_i  |
{\rm e}^{-i\vb.\vr}| \widetilde{\phi}_j \ket \right)
\end{eqnarray}
A similar expression has been used for Ultra-Soft Pseudopotentials by Ferretti
{\it et al}~\cite{Ferretti07}. The first term is computed in the Fourier basis,
whereas the second is computed in the radial grid. We use the expansion
of ${\rm e}^{-i\vb.\vr}$ in spherical harmonics  and bessel functions
in order to compute its expectation value over atomic wavefunctions.

The overlap matrix $M_{m,n}^{(\vk,\vb)}$ is used to minimize the spread and
localize the WFs. This is done with the publicly available Wannier90
code~\cite{mostofi07}.

The code is tested on $t_{2g}$ MLWF for SrVO3 using only the three
KS $t_{2g}$ bands. The spread of one of the WFs for a
$8\times8\times8$ k-point grid centered on the $\Gamma$ point with our
implementation is 6.96 a.u$^2$. In excellent agreement with the FLAPW result
(6.96 a.u$^2$ in Ref.~\onlinecite{Lechermann06}).

\acknowledgements
We are grateful to O.K.~Andersen, F.~Aryasetiawan, S.~Biermann, T.~Miyake,
A.~ Poteryaev, L.~Pourovskii, J.~Tomczak and M.~Torrent for collaborations and
discussions related to the subject of this article. We acknowledge support from
CNRS, Ecole Polytechnique, the Agence Nationale de la Recherche
(under contract {\it ETSF}\,) and the Deutsche Forschungsgemeinschaft (DFG) within
the scope of the SFB 668.

{\bf Note added:}
As the writing of this work was being completed, we became aware of the related work of
Korotin {\it et al.} (arXiv:0801.3500) also reporting on an implementation of LDA+DMFT in a
pseudopotential plane-wave framework.

\bibliographystyle{apsrev}
\bibliography{biblb,bibag,bibextra}

\begin{thebibliography}{55}
\expandafter\ifx\csname natexlab\endcsname\relax\def\natexlab#1{#1}\fi
\expandafter\ifx\csname bibnamefont\endcsname\relax
  \def\bibnamefont#1{#1}\fi
\expandafter\ifx\csname bibfnamefont\endcsname\relax
  \def\bibfnamefont#1{#1}\fi
\expandafter\ifx\csname citenamefont\endcsname\relax
  \def\citenamefont#1{#1}\fi
\expandafter\ifx\csname url\endcsname\relax
  \def\url#1{\texttt{#1}}\fi
\expandafter\ifx\csname urlprefix\endcsname\relax\def\urlprefix{URL }\fi
\providecommand{\bibinfo}[2]{#2}
\providecommand{\eprint}[2][]{\url{#2}}

\bibitem[{\citenamefont{Anisimov et~al.}(2005)\citenamefont{Anisimov, Kondakov,
  Kozhevnikov, Nekrasov, Pchelkina, Allen, Mo, Kim, Metcalf, Suga
  et~al.}}]{Anisimov05}
\bibinfo{author}{\bibfnamefont{V.~I.} \bibnamefont{Anisimov}},
  \bibinfo{author}{\bibfnamefont{D.~E.} \bibnamefont{Kondakov}},
  \bibinfo{author}{\bibfnamefont{A.~V.} \bibnamefont{Kozhevnikov}},
  \bibinfo{author}{\bibfnamefont{I.~A.} \bibnamefont{Nekrasov}},
  \bibinfo{author}{\bibfnamefont{Z.~V.} \bibnamefont{Pchelkina}},
  \bibinfo{author}{\bibfnamefont{J.~W.} \bibnamefont{Allen}},
  \bibinfo{author}{\bibfnamefont{S.-K.} \bibnamefont{Mo}},
  \bibinfo{author}{\bibfnamefont{H.-D.} \bibnamefont{Kim}},
  \bibinfo{author}{\bibfnamefont{P.}~\bibnamefont{Metcalf}},
  \bibinfo{author}{\bibfnamefont{S.}~\bibnamefont{Suga}}, \bibnamefont{et~al.},
  \bibinfo{journal}{Phys. Rev. B} \textbf{\bibinfo{volume}{71}},
  \bibinfo{pages}{125119} (\bibinfo{year}{2005}).

\bibitem[{\citenamefont{Lechermann et~al.}(2006)\citenamefont{Lechermann,
  Georges, Poteryaev, Biermann, Posternak, Yamasaki, and
  Andersen}}]{Lechermann06}
\bibinfo{author}{\bibfnamefont{F.}~\bibnamefont{Lechermann}},
  \bibinfo{author}{\bibfnamefont{A.}~\bibnamefont{Georges}},
  \bibinfo{author}{\bibfnamefont{A.}~\bibnamefont{Poteryaev}},
  \bibinfo{author}{\bibfnamefont{S.}~\bibnamefont{Biermann}},
  \bibinfo{author}{\bibfnamefont{M.}~\bibnamefont{Posternak}},
  \bibinfo{author}{\bibfnamefont{A.}~\bibnamefont{Yamasaki}}, \bibnamefont{and}
  \bibinfo{author}{\bibfnamefont{O.~K.} \bibnamefont{Andersen}},
  \bibinfo{journal}{\prb} \textbf{\bibinfo{volume}{74}}, \bibinfo{eid}{125120}
  (\bibinfo{year}{2006}).

\bibitem[{\citenamefont{{Anisimov} et~al.}(1997)\citenamefont{{Anisimov},
  {Poteryaev}, {Korotin}, {Anokhin}, and {Kotliar}}}]{Anisimov_lda+dmft_1997}
\bibinfo{author}{\bibfnamefont{V.~I.} \bibnamefont{{Anisimov}}},
  \bibinfo{author}{\bibfnamefont{A.~I.} \bibnamefont{{Poteryaev}}},
  \bibinfo{author}{\bibfnamefont{M.~A.} \bibnamefont{{Korotin}}},
  \bibinfo{author}{\bibfnamefont{A.~O.} \bibnamefont{{Anokhin}}},
  \bibnamefont{and}
  \bibinfo{author}{\bibfnamefont{G.}~\bibnamefont{{Kotliar}}},
  \bibinfo{journal}{J. Phys. Cond. Matter} \textbf{\bibinfo{volume}{9}},
  \bibinfo{pages}{7359} (\bibinfo{year}{1997}).

\bibitem[{\citenamefont{{Lichtenstein} and
  {Katsnelson}}(1998)}]{Lichtenstein_lda+dmft_1998}
\bibinfo{author}{\bibfnamefont{A.~I.} \bibnamefont{{Lichtenstein}}}
  \bibnamefont{and} \bibinfo{author}{\bibfnamefont{M.~I.}
  \bibnamefont{{Katsnelson}}}, \bibinfo{journal}{Phys. Rev. B}
  \textbf{\bibinfo{volume}{57}}, \bibinfo{pages}{6884} (\bibinfo{year}{1998}).

\bibitem[{\citenamefont{{Savrasov} et~al.}(2001)\citenamefont{{Savrasov},
  {Kotliar}, and Abrahams}}]{Savrasov_Kotliar_pu_nature_2001}
\bibinfo{author}{\bibfnamefont{S.~Y.} \bibnamefont{{Savrasov}}},
  \bibinfo{author}{\bibfnamefont{G.}~\bibnamefont{{Kotliar}}},
  \bibnamefont{and} \bibinfo{author}{\bibfnamefont{E.}~\bibnamefont{Abrahams}},
  \bibinfo{journal}{Nature} \textbf{\bibinfo{volume}{410}},
  \bibinfo{pages}{793} (\bibinfo{year}{2001}).

\bibitem[{\citenamefont{{Andersen}}(1975)}]{Andersen_lmto_1975_prb}
\bibinfo{author}{\bibfnamefont{O.~K.} \bibnamefont{{Andersen}}},
  \bibinfo{journal}{Phys. Rev. B} \textbf{\bibinfo{volume}{12}},
  \bibinfo{pages}{3060} (\bibinfo{year}{1975}).

\bibitem[{\citenamefont{Pavarini
  et~al.}(2004{\natexlab{a}})\citenamefont{Pavarini, Biermann, Poteryaev,
  Lichtenstein, Georges, and Andersen}}]{Pavarini04}
\bibinfo{author}{\bibfnamefont{E.}~\bibnamefont{Pavarini}},
  \bibinfo{author}{\bibfnamefont{S.}~\bibnamefont{Biermann}},
  \bibinfo{author}{\bibfnamefont{A.}~\bibnamefont{Poteryaev}},
  \bibinfo{author}{\bibfnamefont{A.~I.} \bibnamefont{Lichtenstein}},
  \bibinfo{author}{\bibfnamefont{A.}~\bibnamefont{Georges}}, \bibnamefont{and}
  \bibinfo{author}{\bibfnamefont{O.~K.} \bibnamefont{Andersen}},
  \bibinfo{journal}{Phys. Rev. Lett.} \textbf{\bibinfo{volume}{92}},
  \bibinfo{pages}{176403} (\bibinfo{year}{2004}{\natexlab{a}}).

\bibitem[{\citenamefont{Andersen and Saha-Dasgupta}(2000)}]{Andersen_NMTO_00}
\bibinfo{author}{\bibfnamefont{O.~K.} \bibnamefont{Andersen}} \bibnamefont{and}
  \bibinfo{author}{\bibfnamefont{T.}~\bibnamefont{Saha-Dasgupta}},
  \bibinfo{journal}{Phys. Rev. B} \textbf{\bibinfo{volume}{62}},
  \bibinfo{pages}{16219} (\bibinfo{year}{2000}).

\bibitem[{\citenamefont{des Cloizeaux}(1963)}]{Cloizeaux63}
\bibinfo{author}{\bibfnamefont{J.}~\bibnamefont{des Cloizeaux}},
  \bibinfo{journal}{Phys. Rev.} \textbf{\bibinfo{volume}{129}},
  \bibinfo{pages}{554} (\bibinfo{year}{1963}).

\bibitem[{\citenamefont{Ku et~al.}(2002)\citenamefont{Ku, Rosner, Pickett, and
  Scalettar}}]{ku02}
\bibinfo{author}{\bibfnamefont{W.}~\bibnamefont{Ku}},
  \bibinfo{author}{\bibfnamefont{H.}~\bibnamefont{Rosner}},
  \bibinfo{author}{\bibfnamefont{W.~E.} \bibnamefont{Pickett}},
  \bibnamefont{and} \bibinfo{author}{\bibfnamefont{R.~T.}
  \bibnamefont{Scalettar}}, \bibinfo{journal}{Phys. Rev. Lett.}
  \textbf{\bibinfo{volume}{89}}, \bibinfo{pages}{167204}
  (\bibinfo{year}{2002}).

\bibitem[{\citenamefont{Marzari and Vanderbilt}(1997)}]{Marzari97}
\bibinfo{author}{\bibfnamefont{N.}~\bibnamefont{Marzari}} \bibnamefont{and}
  \bibinfo{author}{\bibfnamefont{D.}~\bibnamefont{Vanderbilt}},
  \bibinfo{journal}{Phys. Rev. B} \textbf{\bibinfo{volume}{56}},
  \bibinfo{pages}{12847} (\bibinfo{year}{1997}).

\bibitem[{\citenamefont{Imada et~al.}(1998)\citenamefont{Imada, Fujimori, and
  Tokura}}]{imada_mit_review}
\bibinfo{author}{\bibfnamefont{M.}~\bibnamefont{Imada}},
  \bibinfo{author}{\bibfnamefont{A.}~\bibnamefont{Fujimori}}, \bibnamefont{and}
  \bibinfo{author}{\bibfnamefont{Y.}~\bibnamefont{Tokura}},
  \bibinfo{journal}{Rev. Mod. Phys.} \textbf{\bibinfo{volume}{70}},
  \bibinfo{pages}{1039} (\bibinfo{year}{1998}).

\bibitem[{\citenamefont{Fujimori et~al.}(1992)\citenamefont{Fujimori, Hase,
  Namatame, Fujishima, Tokura, Eisaki, Uchida, Takegahara, and
  de~Groot}}]{fujimori_pes_oxides}
\bibinfo{author}{\bibfnamefont{A.}~\bibnamefont{Fujimori}},
  \bibinfo{author}{\bibfnamefont{I.}~\bibnamefont{Hase}},
  \bibinfo{author}{\bibfnamefont{H.}~\bibnamefont{Namatame}},
  \bibinfo{author}{\bibfnamefont{Y.}~\bibnamefont{Fujishima}},
  \bibinfo{author}{\bibfnamefont{Y.}~\bibnamefont{Tokura}},
  \bibinfo{author}{\bibfnamefont{H.}~\bibnamefont{Eisaki}},
  \bibinfo{author}{\bibfnamefont{S.}~\bibnamefont{Uchida}},
  \bibinfo{author}{\bibfnamefont{K.}~\bibnamefont{Takegahara}},
  \bibnamefont{and} \bibinfo{author}{\bibfnamefont{F.~M.~F.}
  \bibnamefont{de~Groot}}, \bibinfo{journal}{Phys. Rev. Lett.}
  \textbf{\bibinfo{volume}{69}}, \bibinfo{pages}{1796} (\bibinfo{year}{1992}).

\bibitem[{\citenamefont{Maiti et~al.}(2001)\citenamefont{Maiti, Sarma,
  Rozenberg, Inoue, Makino, Goto, Pedio, and Cimino}}]{maiti_2001}
\bibinfo{author}{\bibfnamefont{K.}~\bibnamefont{Maiti}},
  \bibinfo{author}{\bibfnamefont{D.~D.} \bibnamefont{Sarma}},
  \bibinfo{author}{\bibfnamefont{M.}~\bibnamefont{Rozenberg}},
  \bibinfo{author}{\bibfnamefont{I.}~\bibnamefont{Inoue}},
  \bibinfo{author}{\bibfnamefont{H.}~\bibnamefont{Makino}},
  \bibinfo{author}{\bibfnamefont{O.}~\bibnamefont{Goto}},
  \bibinfo{author}{\bibfnamefont{M.}~\bibnamefont{Pedio}}, \bibnamefont{and}
  \bibinfo{author}{\bibfnamefont{R.}~\bibnamefont{Cimino}},
  \bibinfo{journal}{Europhys. Lett.} \textbf{\bibinfo{volume}{55}},
  \bibinfo{pages}{246} (\bibinfo{year}{2001}).

\bibitem[{\citenamefont{Maiti}(1997)}]{maiti_phd}
\bibinfo{author}{\bibfnamefont{K.}~\bibnamefont{Maiti}}, Ph.D. thesis,
  \bibinfo{school}{IISC, Bangalore} (\bibinfo{year}{1997}).

\bibitem[{\citenamefont{{Inoue} et~al.}(1995)\citenamefont{{Inoue}, {Hase},
  {Aiura}, {Fujimori}, {Haruyama}, {Maruyama}, and
  {Nishihara}}}]{inoue_casrvo3_1995_prl}
\bibinfo{author}{\bibfnamefont{I.~H.} \bibnamefont{{Inoue}}},
  \bibinfo{author}{\bibfnamefont{I.}~\bibnamefont{{Hase}}},
  \bibinfo{author}{\bibfnamefont{Y.}~\bibnamefont{{Aiura}}},
  \bibinfo{author}{\bibfnamefont{A.}~\bibnamefont{{Fujimori}}},
  \bibinfo{author}{\bibfnamefont{Y.}~\bibnamefont{{Haruyama}}},
  \bibinfo{author}{\bibfnamefont{T.}~\bibnamefont{{Maruyama}}},
  \bibnamefont{and}
  \bibinfo{author}{\bibfnamefont{Y.}~\bibnamefont{{Nishihara}}},
  \bibinfo{journal}{Phys. Rev. Lett.} \textbf{\bibinfo{volume}{74}},
  \bibinfo{pages}{2539} (\bibinfo{year}{1995}).

\bibitem[{\citenamefont{Sekiyama et~al.}(2004)\citenamefont{Sekiyama, Fujiwara,
  Imada, Suga, Eisaki, Uchida, Takegahara, Harima, Saitoh, Nekrasov
  et~al.}}]{sek04}
\bibinfo{author}{\bibfnamefont{A.}~\bibnamefont{Sekiyama}},
  \bibinfo{author}{\bibfnamefont{H.}~\bibnamefont{Fujiwara}},
  \bibinfo{author}{\bibfnamefont{S.}~\bibnamefont{Imada}},
  \bibinfo{author}{\bibfnamefont{S.}~\bibnamefont{Suga}},
  \bibinfo{author}{\bibfnamefont{H.}~\bibnamefont{Eisaki}},
  \bibinfo{author}{\bibfnamefont{S.~I.} \bibnamefont{Uchida}},
  \bibinfo{author}{\bibfnamefont{K.}~\bibnamefont{Takegahara}},
  \bibinfo{author}{\bibfnamefont{H.}~\bibnamefont{Harima}},
  \bibinfo{author}{\bibfnamefont{Y.}~\bibnamefont{Saitoh}},
  \bibinfo{author}{\bibfnamefont{I.~A.} \bibnamefont{Nekrasov}},
  \bibnamefont{et~al.}, \bibinfo{journal}{Phys. Rev. Lett.}
  \textbf{\bibinfo{volume}{93}}, \bibinfo{pages}{156402}
  (\bibinfo{year}{2004}).

\bibitem[{\citenamefont{Liebsch}(2003)}]{lie03}
\bibinfo{author}{\bibfnamefont{A.}~\bibnamefont{Liebsch}},
  \bibinfo{journal}{Phys. Rev. Lett.} \textbf{\bibinfo{volume}{90}},
  \bibinfo{pages}{096401} (\bibinfo{year}{2003}).

\bibitem[{\citenamefont{Nekrasov et~al.}(2005)\citenamefont{Nekrasov, Keller,
  Kondakov, , Kozhevnikov, Pruschke, Held, Vollhardt, and Anisimov}}]{nek05}
\bibinfo{author}{\bibfnamefont{I.~A.} \bibnamefont{Nekrasov}},
  \bibinfo{author}{\bibfnamefont{G.}~\bibnamefont{Keller}},
  \bibinfo{author}{\bibfnamefont{D.~E.} \bibnamefont{Kondakov}}, ,
  \bibinfo{author}{\bibfnamefont{A.~V.} \bibnamefont{Kozhevnikov}},
  \bibinfo{author}{\bibfnamefont{T.}~\bibnamefont{Pruschke}},
  \bibinfo{author}{\bibfnamefont{K.}~\bibnamefont{Held}},
  \bibinfo{author}{\bibfnamefont{D.}~\bibnamefont{Vollhardt}},
  \bibnamefont{and} \bibinfo{author}{\bibfnamefont{V.~I.}
  \bibnamefont{Anisimov}}, \bibinfo{journal}{Phys. Rev. B}
  \textbf{\bibinfo{volume}{72}}, \bibinfo{pages}{155106}
  (\bibinfo{year}{2005}).

\bibitem[{\citenamefont{Pavarini
  et~al.}(2004{\natexlab{b}})\citenamefont{Pavarini, Biermann, Poteryaev,
  Lichtenstein, Georges, and Andersen}}]{pav04}
\bibinfo{author}{\bibfnamefont{E.}~\bibnamefont{Pavarini}},
  \bibinfo{author}{\bibfnamefont{S.}~\bibnamefont{Biermann}},
  \bibinfo{author}{\bibfnamefont{A.}~\bibnamefont{Poteryaev}},
  \bibinfo{author}{\bibfnamefont{A.~I.} \bibnamefont{Lichtenstein}},
  \bibinfo{author}{\bibfnamefont{A.}~\bibnamefont{Georges}}, \bibnamefont{and}
  \bibinfo{author}{\bibfnamefont{O.~K.} \bibnamefont{Andersen}},
  \bibinfo{journal}{Phys. Rev. Lett.} \textbf{\bibinfo{volume}{92}},
  \bibinfo{pages}{176403} (\bibinfo{year}{2004}{\natexlab{b}}).

\bibitem[{\citenamefont{Yoshida et~al.}(2005)\citenamefont{Yoshida, Tanaka,
  Yagi, Ino, Eisaki, Fujimori, and Shen}}]{yos05}
\bibinfo{author}{\bibfnamefont{T.}~\bibnamefont{Yoshida}},
  \bibinfo{author}{\bibfnamefont{K.}~\bibnamefont{Tanaka}},
  \bibinfo{author}{\bibfnamefont{H.}~\bibnamefont{Yagi}},
  \bibinfo{author}{\bibfnamefont{A.}~\bibnamefont{Ino}},
  \bibinfo{author}{\bibfnamefont{H.}~\bibnamefont{Eisaki}},
  \bibinfo{author}{\bibfnamefont{A.}~\bibnamefont{Fujimori}}, \bibnamefont{and}
  \bibinfo{author}{\bibfnamefont{Z.-X.} \bibnamefont{Shen}},
  \bibinfo{journal}{Phys. Rev. Lett.} \textbf{\bibinfo{volume}{95}},
  \bibinfo{pages}{146404} (\bibinfo{year}{2005}).

\bibitem[{\citenamefont{Wadati et~al.}(2006)\citenamefont{Wadati, Yoshida,
  Chikamatsu, Kumigashira, Oshima, Eisaki, Shen, Mizokawa, and
  Fujimori}}]{wad06}
\bibinfo{author}{\bibfnamefont{H.}~\bibnamefont{Wadati}},
  \bibinfo{author}{\bibfnamefont{T.}~\bibnamefont{Yoshida}},
  \bibinfo{author}{\bibfnamefont{A.}~\bibnamefont{Chikamatsu}},
  \bibinfo{author}{\bibfnamefont{H.}~\bibnamefont{Kumigashira}},
  \bibinfo{author}{\bibfnamefont{M.}~\bibnamefont{Oshima}},
  \bibinfo{author}{\bibfnamefont{H.}~\bibnamefont{Eisaki}},
  \bibinfo{author}{\bibfnamefont{Z.~X.} \bibnamefont{Shen}},
  \bibinfo{author}{\bibfnamefont{T.}~\bibnamefont{Mizokawa}}, \bibnamefont{and}
  \bibinfo{author}{\bibfnamefont{A.}~\bibnamefont{Fujimori}},
  \bibinfo{journal}{cond-mat/0603642}  (\bibinfo{year}{2006}).

\bibitem[{\citenamefont{Solovyev}(2006)}]{sol06}
\bibinfo{author}{\bibfnamefont{I.~V.} \bibnamefont{Solovyev}},
  \bibinfo{journal}{Phys. Rev. B} \textbf{\bibinfo{volume}{73}},
  \bibinfo{pages}{155117} (\bibinfo{year}{2006}).

\bibitem[{\citenamefont{Nekrasov et~al.}(2006)\citenamefont{Nekrasov, Held,
  Keller, Kondakov, Pruschke, Kollar, Andersen, Anisimov, and
  Vollhardt}}]{nek06}
\bibinfo{author}{\bibfnamefont{I.~A.} \bibnamefont{Nekrasov}},
  \bibinfo{author}{\bibfnamefont{K.}~\bibnamefont{Held}},
  \bibinfo{author}{\bibfnamefont{G.}~\bibnamefont{Keller}},
  \bibinfo{author}{\bibfnamefont{D.~E.} \bibnamefont{Kondakov}},
  \bibinfo{author}{\bibfnamefont{T.}~\bibnamefont{Pruschke}},
  \bibinfo{author}{\bibfnamefont{M.}~\bibnamefont{Kollar}},
  \bibinfo{author}{\bibfnamefont{O.~K.} \bibnamefont{Andersen}},
  \bibinfo{author}{\bibfnamefont{V.~I.} \bibnamefont{Anisimov}},
  \bibnamefont{and}
  \bibinfo{author}{\bibfnamefont{D.}~\bibnamefont{Vollhardt}},
  \bibinfo{journal}{Phys. Rev. B} \textbf{\bibinfo{volume}{73}},
  \bibinfo{pages}{155112} (\bibinfo{year}{2006}).

\bibitem[{\citenamefont{Eguchi et~al.}(2006)\citenamefont{Eguchi, Kiss, Tsuda,
  Shimojima, Mizokami, Yokoya, Chainani, Shin, Inoue, Togashi et~al.}}]{egu06}
\bibinfo{author}{\bibfnamefont{R.}~\bibnamefont{Eguchi}},
  \bibinfo{author}{\bibfnamefont{T.}~\bibnamefont{Kiss}},
  \bibinfo{author}{\bibfnamefont{S.}~\bibnamefont{Tsuda}},
  \bibinfo{author}{\bibfnamefont{T.}~\bibnamefont{Shimojima}},
  \bibinfo{author}{\bibfnamefont{T.}~\bibnamefont{Mizokami}},
  \bibinfo{author}{\bibfnamefont{T.}~\bibnamefont{Yokoya}},
  \bibinfo{author}{\bibfnamefont{A.}~\bibnamefont{Chainani}},
  \bibinfo{author}{\bibfnamefont{S.}~\bibnamefont{Shin}},
  \bibinfo{author}{\bibfnamefont{I.~H.} \bibnamefont{Inoue}},
  \bibinfo{author}{\bibfnamefont{T.}~\bibnamefont{Togashi}},
  \bibnamefont{et~al.}, \bibinfo{journal}{Phys. Rev. Lett.}
  \textbf{\bibinfo{volume}{96}}, \bibinfo{pages}{076402}
  (\bibinfo{year}{2006}).

\bibitem[{\citenamefont{{Tackett} et~al.}(2001)\citenamefont{{Tackett},
  {Holzwarth}, and {Matthews}}}]{atompaw}
\bibinfo{author}{\bibfnamefont{A.~R.} \bibnamefont{{Tackett}}},
  \bibinfo{author}{\bibfnamefont{N.~A.~W.} \bibnamefont{{Holzwarth}}},
  \bibnamefont{and} \bibinfo{author}{\bibfnamefont{G.~E.}
  \bibnamefont{{Matthews}}}, \bibinfo{journal}{Comp. Phys. Com.}
  \textbf{\bibinfo{volume}{135}} (\bibinfo{year}{2001}).

\bibitem[{\citenamefont{{Holzwarth} et~al.}(2007)\citenamefont{{Holzwarth},
  Torrent, and Jollet}}]{AtompawAbinit}
\bibinfo{author}{\bibfnamefont{N.~A.~W.} \bibnamefont{{Holzwarth}}},
  \bibinfo{author}{\bibfnamefont{M.}~\bibnamefont{Torrent}}, \bibnamefont{and}
  \bibinfo{author}{\bibfnamefont{F.}~\bibnamefont{Jollet}},
  \bibinfo{journal}{ATOMPAW (http://pwpaw.wfu.edu/)}  (\bibinfo{year}{2007}).

\bibitem[{\citenamefont{Gonze et~al.}(2002)\citenamefont{Gonze, Beuken,
  Caracas, Detraux, Fuchs, Rignanese, Sindic, Verstraete, Zerah, Jollet
  et~al.}}]{abinit1}
\bibinfo{author}{\bibfnamefont{X.}~\bibnamefont{Gonze}},
  \bibinfo{author}{\bibfnamefont{J.-M.} \bibnamefont{Beuken}},
  \bibinfo{author}{\bibfnamefont{R.}~\bibnamefont{Caracas}},
  \bibinfo{author}{\bibfnamefont{F.}~\bibnamefont{Detraux}},
  \bibinfo{author}{\bibfnamefont{M.}~\bibnamefont{Fuchs}},
  \bibinfo{author}{\bibfnamefont{G.-M.} \bibnamefont{Rignanese}},
  \bibinfo{author}{\bibfnamefont{L.}~\bibnamefont{Sindic}},
  \bibinfo{author}{\bibfnamefont{M.}~\bibnamefont{Verstraete}},
  \bibinfo{author}{\bibfnamefont{G.}~\bibnamefont{Zerah}},
  \bibinfo{author}{\bibfnamefont{F.}~\bibnamefont{Jollet}},
  \bibnamefont{et~al.}, \bibinfo{journal}{Comp. Mat. Science}
  \textbf{\bibinfo{volume}{25}}, \bibinfo{pages}{478} (\bibinfo{year}{2002}).

\bibitem[{\citenamefont{Torrent et~al.}(2007 (in press))\citenamefont{Torrent,
  jollet, Bottin, Zerah, and Gonze}}]{torrent07}
\bibinfo{author}{\bibfnamefont{M.}~\bibnamefont{Torrent}},
  \bibinfo{author}{\bibfnamefont{F.}~\bibnamefont{jollet}},
  \bibinfo{author}{\bibfnamefont{F.}~\bibnamefont{Bottin}},
  \bibinfo{author}{\bibfnamefont{G.}~\bibnamefont{Zerah}}, \bibnamefont{and}
  \bibinfo{author}{\bibfnamefont{X.}~\bibnamefont{Gonze}},
  \bibinfo{journal}{Comput. Mater. Sci.}  (\bibinfo{year}{2007 (in press)}),
  \eprint{doi: 10.1016/j.commatsci.2007.07.020}.

\bibitem[{\citenamefont{Fr\'esard and Kotliar}(1997)}]{Fresard97}
\bibinfo{author}{\bibfnamefont{R.}~\bibnamefont{Fr\'esard}} \bibnamefont{and}
  \bibinfo{author}{\bibfnamefont{G.}~\bibnamefont{Kotliar}},
  \bibinfo{journal}{Phys. Rev. B} \textbf{\bibinfo{volume}{56}},
  \bibinfo{pages}{12909} (\bibinfo{year}{1997}).

\bibitem[{\citenamefont{Maiti et~al.}(2005)\citenamefont{Maiti, Manju, Ray,
  Mahadevan, and I.~H.~Inoue}}]{mai05}
\bibinfo{author}{\bibfnamefont{K.}~\bibnamefont{Maiti}},
  \bibinfo{author}{\bibfnamefont{U.}~\bibnamefont{Manju}},
  \bibinfo{author}{\bibfnamefont{S.}~\bibnamefont{Ray}},
  \bibinfo{author}{\bibfnamefont{P.}~\bibnamefont{Mahadevan}},
  \bibnamefont{and} \bibinfo{author}{\bibfnamefont{D.~D.~S.}
  \bibnamefont{I.~H.~Inoue}, \bibfnamefont{C.~Carbone}},
  \bibinfo{journal}{cond-mat/0509643}  (\bibinfo{year}{2005}).

\bibitem[{\citenamefont{Sparks and Komoto}(1967)}]{Spa67}
\bibinfo{author}{\bibfnamefont{J.~T.} \bibnamefont{Sparks}} \bibnamefont{and}
  \bibinfo{author}{\bibfnamefont{T.}~\bibnamefont{Komoto}},
  \bibinfo{journal}{Phys. Lett.} \textbf{\bibinfo{volume}{25}},
  \bibinfo{pages}{398} (\bibinfo{year}{1967}).

\bibitem[{\citenamefont{White and Mott}(1971)}]{Whi71}
\bibinfo{author}{\bibfnamefont{R.~M.} \bibnamefont{White}} \bibnamefont{and}
  \bibinfo{author}{\bibfnamefont{N.~F.} \bibnamefont{Mott}},
  \bibinfo{journal}{Philos. Mag.} \textbf{\bibinfo{volume}{24}},
  \bibinfo{pages}{845} (\bibinfo{year}{1971}).

\bibitem[{\citenamefont{Ohtani et~al.}(1970)\citenamefont{Ohtani, Kosuge, and
  Kachi}}]{Oht70}
\bibinfo{author}{\bibfnamefont{T.}~\bibnamefont{Ohtani}},
  \bibinfo{author}{\bibfnamefont{K.}~\bibnamefont{Kosuge}}, \bibnamefont{and}
  \bibinfo{author}{\bibfnamefont{S.}~\bibnamefont{Kachi}}, \bibinfo{journal}{J.
  Phys. Soc. Jpn.} \textbf{\bibinfo{volume}{28}}, \bibinfo{pages}{1588}
  (\bibinfo{year}{1970}).

\bibitem[{\citenamefont{Benoit}(1955)}]{Ben55}
\bibinfo{author}{\bibfnamefont{R.}~\bibnamefont{Benoit}},
  \bibinfo{journal}{J.~Chem.~Phys.} \textbf{\bibinfo{volume}{52}},
  \bibinfo{pages}{119} (\bibinfo{year}{1955}).

\bibitem[{\citenamefont{Trahan et~al.}(1970)\citenamefont{Trahan, Goodrich, and
  Watkins}}]{Tra70}
\bibinfo{author}{\bibfnamefont{J.}~\bibnamefont{Trahan}},
  \bibinfo{author}{\bibfnamefont{R.~G.} \bibnamefont{Goodrich}},
  \bibnamefont{and} \bibinfo{author}{\bibfnamefont{S.~F.}
  \bibnamefont{Watkins}}, \bibinfo{journal}{Phys. Rev. B}
  \textbf{\bibinfo{volume}{2}}, \bibinfo{pages}{2859} (\bibinfo{year}{1970}).

\bibitem[{\citenamefont{Zaanen et~al.}(1985)\citenamefont{Zaanen, Sawatzky, and
  Allen}}]{zsa_1985_prl}
\bibinfo{author}{\bibfnamefont{J.}~\bibnamefont{Zaanen}},
  \bibinfo{author}{\bibfnamefont{G.~A.} \bibnamefont{Sawatzky}},
  \bibnamefont{and} \bibinfo{author}{\bibfnamefont{J.~W.} \bibnamefont{Allen}},
  \bibinfo{journal}{Phys. Rev. Lett.} \textbf{\bibinfo{volume}{55}},
  \bibinfo{pages}{418} (\bibinfo{year}{1985}).

\bibitem[{\citenamefont{Mattheiss}(1974)}]{mat74}
\bibinfo{author}{\bibfnamefont{L.}~\bibnamefont{Mattheiss}},
  \bibinfo{journal}{Phys. Rev. B} \textbf{\bibinfo{volume}{10}},
  \bibinfo{pages}{995} (\bibinfo{year}{1974}).

\bibitem[{\citenamefont{Meyer et~al.}(unpublished)\citenamefont{Meyer,
  Els\"{a}sser, Lechermann, and F\"{a}hnle}}]{mbpp_code}
\bibinfo{author}{\bibfnamefont{B.}~\bibnamefont{Meyer}},
  \bibinfo{author}{\bibfnamefont{C.}~\bibnamefont{Els\"{a}sser}},
  \bibinfo{author}{\bibfnamefont{F.}~\bibnamefont{Lechermann}},
  \bibnamefont{and}
  \bibinfo{author}{\bibfnamefont{M.}~\bibnamefont{F\"{a}hnle}},
  \emph{\bibinfo{title}{FORTRAN 90 Program for Mixed-Basis-Pseudopotential
  Calculations for Crystals}}, \bibinfo{organization}{Max-Planck-Institut
  f\"{u}r Metallforschung, Stuttgart} (\bibinfo{year}{unpublished}).

\bibitem[{\citenamefont{H\"{u}fner et~al.}(1985)\citenamefont{H\"{u}fner,
  Riesterer, and Hulliger}}]{huf85}
\bibinfo{author}{\bibfnamefont{S.}~\bibnamefont{H\"{u}fner}},
  \bibinfo{author}{\bibfnamefont{T.}~\bibnamefont{Riesterer}},
  \bibnamefont{and} \bibinfo{author}{\bibfnamefont{F.}~\bibnamefont{Hulliger}},
  \bibinfo{journal}{Solid State Comm.} \textbf{\bibinfo{volume}{54}},
  \bibinfo{pages}{689} (\bibinfo{year}{1985}).

\bibitem[{\citenamefont{Fujimori et~al.}(1988)\citenamefont{Fujimori, Terakura,
  Taniguchi, Ogawa, Suga, Matoba, and Anzai}}]{fuj88}
\bibinfo{author}{\bibfnamefont{A.}~\bibnamefont{Fujimori}},
  \bibinfo{author}{\bibfnamefont{K.}~\bibnamefont{Terakura}},
  \bibinfo{author}{\bibfnamefont{M.}~\bibnamefont{Taniguchi}},
  \bibinfo{author}{\bibfnamefont{S.}~\bibnamefont{Ogawa}},
  \bibinfo{author}{\bibfnamefont{S.}~\bibnamefont{Suga}},
  \bibinfo{author}{\bibfnamefont{M.}~\bibnamefont{Matoba}}, \bibnamefont{and}
  \bibinfo{author}{\bibfnamefont{S.}~\bibnamefont{Anzai}},
  \bibinfo{journal}{Phys. Rev. B} \textbf{\bibinfo{volume}{37}},
  \bibinfo{pages}{3109} (\bibinfo{year}{1988}).

\bibitem[{\citenamefont{{Anisimov} et~al.}(1991)\citenamefont{{Anisimov},
  {Zaanen}, and {Andersen}}}]{Anisimov91}
\bibinfo{author}{\bibfnamefont{V.~I.} \bibnamefont{{Anisimov}}},
  \bibinfo{author}{\bibfnamefont{J.}~\bibnamefont{{Zaanen}}}, \bibnamefont{and}
  \bibinfo{author}{\bibfnamefont{O.~K.} \bibnamefont{{Andersen}}},
  \bibinfo{journal}{\prb} \textbf{\bibinfo{volume}{44}}, \bibinfo{pages}{943}
  (\bibinfo{year}{1991}).

\bibitem[{\citenamefont{Nakamura et~al.}(1994)\citenamefont{Nakamura, Sekiyama,
  Namatame, Kino, and Fujimori}}]{nak94b}
\bibinfo{author}{\bibfnamefont{M.}~\bibnamefont{Nakamura}},
  \bibinfo{author}{\bibfnamefont{A.}~\bibnamefont{Sekiyama}},
  \bibinfo{author}{\bibfnamefont{H.}~\bibnamefont{Namatame}},
  \bibinfo{author}{\bibfnamefont{H.}~\bibnamefont{Kino}}, \bibnamefont{and}
  \bibinfo{author}{\bibfnamefont{A.}~\bibnamefont{Fujimori}},
  \bibinfo{journal}{Phys. Rev. Lett.} \textbf{\bibinfo{volume}{73}},
  \bibinfo{pages}{2891} (\bibinfo{year}{1994}).

\bibitem[{\citenamefont{Usuda and Hamada}(2000)}]{usu00}
\bibinfo{author}{\bibfnamefont{M.}~\bibnamefont{Usuda}} \bibnamefont{and}
  \bibinfo{author}{\bibfnamefont{N.}~\bibnamefont{Hamada}},
  \bibinfo{journal}{J. Phys. Chem.} \textbf{\bibinfo{volume}{69}},
  \bibinfo{pages}{744} (\bibinfo{year}{2000}).

\bibitem[{\citenamefont{Krishnakumar et~al.}(2000)\citenamefont{Krishnakumar,
  Shanthi, Mahadevan, and Sarma}}]{Kri00}
\bibinfo{author}{\bibfnamefont{S.~R.} \bibnamefont{Krishnakumar}},
  \bibinfo{author}{\bibfnamefont{N.}~\bibnamefont{Shanthi}},
  \bibinfo{author}{\bibfnamefont{P.}~\bibnamefont{Mahadevan}},
  \bibnamefont{and} \bibinfo{author}{\bibfnamefont{D.~D.} \bibnamefont{Sarma}},
  \bibinfo{journal}{Phys. Rev. B} \textbf{\bibinfo{volume}{61}},
  \bibinfo{pages}{16370} (\bibinfo{year}{2000}).

\bibitem[{\citenamefont{Kune\v{s}
  et~al.}(2007{\natexlab{a}})\citenamefont{Kune\v{s}, Anisimov, Lukoyanov, and
  Vollhardt}}]{kun07a}
\bibinfo{author}{\bibfnamefont{J.}~\bibnamefont{Kune\v{s}}},
  \bibinfo{author}{\bibfnamefont{V.~I.} \bibnamefont{Anisimov}},
  \bibinfo{author}{\bibfnamefont{A.~V.} \bibnamefont{Lukoyanov}},
  \bibnamefont{and}
  \bibinfo{author}{\bibfnamefont{D.}~\bibnamefont{Vollhardt}},
  \bibinfo{journal}{Phys. Rev. B} \textbf{\bibinfo{volume}{75}},
  \bibinfo{pages}{165115} (\bibinfo{year}{2007}{\natexlab{a}}).

\bibitem[{\citenamefont{Kune\v{s}
  et~al.}(2007{\natexlab{b}})\citenamefont{Kune\v{s}, Anisimov, Skornyakov,
  Lukoyanov, and Vollhardt}}]{kun07b}
\bibinfo{author}{\bibfnamefont{J.}~\bibnamefont{Kune\v{s}}},
  \bibinfo{author}{\bibfnamefont{V.~I.} \bibnamefont{Anisimov}},
  \bibinfo{author}{\bibfnamefont{S.~L.} \bibnamefont{Skornyakov}},
  \bibinfo{author}{\bibfnamefont{A.~V.} \bibnamefont{Lukoyanov}},
  \bibnamefont{and}
  \bibinfo{author}{\bibfnamefont{D.}~\bibnamefont{Vollhardt}},
  \bibinfo{journal}{Phys. Rev. Lett.} \textbf{\bibinfo{volume}{99}},
  \bibinfo{pages}{156404} (\bibinfo{year}{2007}{\natexlab{b}}).

\bibitem[{\citenamefont{{Bl{\"o}chl}}(1994)}]{Blochl94}
\bibinfo{author}{\bibfnamefont{P.~E.} \bibnamefont{{Bl{\"o}chl}}},
  \bibinfo{journal}{\prb} \textbf{\bibinfo{volume}{50}}, \bibinfo{pages}{17953}
  (\bibinfo{year}{1994}).

\bibitem[{\citenamefont{Louie et~al.}(1979)\citenamefont{Louie, Ho, and
  Cohen}}]{lou79}
\bibinfo{author}{\bibfnamefont{S.~G.} \bibnamefont{Louie}},
  \bibinfo{author}{\bibfnamefont{K.~M.} \bibnamefont{Ho}}, \bibnamefont{and}
  \bibinfo{author}{\bibfnamefont{M.~L.} \bibnamefont{Cohen}},
  \bibinfo{journal}{Phys. Rev. B} \textbf{\bibinfo{volume}{19}},
  \bibinfo{pages}{1774} (\bibinfo{year}{1979}).

\bibitem[{\citenamefont{{Bengone} et~al.}(2000)\citenamefont{{Bengone},
  {Alouani}, {Bl{\"o}chl}, and {Hugel}}}]{Bengone00}
\bibinfo{author}{\bibfnamefont{O.}~\bibnamefont{{Bengone}}},
  \bibinfo{author}{\bibfnamefont{M.}~\bibnamefont{{Alouani}}},
  \bibinfo{author}{\bibfnamefont{P.}~\bibnamefont{{Bl{\"o}chl}}},
  \bibnamefont{and} \bibinfo{author}{\bibfnamefont{J.}~\bibnamefont{{Hugel}}},
  \bibinfo{journal}{\prb} \textbf{\bibinfo{volume}{62}}, \bibinfo{pages}{16392}
  (\bibinfo{year}{2000}).

\bibitem[{\citenamefont{Amadon et~al.}(2007)\citenamefont{Amadon, Jollet, and
  Torrent}}]{Amadon07b}
\bibinfo{author}{\bibfnamefont{B.}~\bibnamefont{Amadon}},
  \bibinfo{author}{\bibfnamefont{F.}~\bibnamefont{Jollet}}, \bibnamefont{and}
  \bibinfo{author}{\bibfnamefont{M.}~\bibnamefont{Torrent}}
  (\bibinfo{year}{2007}).

\bibitem[{\citenamefont{Kresse and Hafner}(1994)}]{kre94}
\bibinfo{author}{\bibfnamefont{G.}~\bibnamefont{Kresse}} \bibnamefont{and}
  \bibinfo{author}{\bibfnamefont{J.}~\bibnamefont{Hafner}},
  \bibinfo{journal}{J. Phys.: Condens. Matter} \textbf{\bibinfo{volume}{6}},
  \bibinfo{pages}{8245} (\bibinfo{year}{1994}).

\bibitem[{\citenamefont{{Czy{\.z}yk} and {Sawatzky}}(1994)}]{Czyzyk94}
\bibinfo{author}{\bibfnamefont{M.~T.} \bibnamefont{{Czy{\.z}yk}}}
  \bibnamefont{and} \bibinfo{author}{\bibfnamefont{G.~A.}
  \bibnamefont{{Sawatzky}}}, \bibinfo{journal}{\prb}
  \textbf{\bibinfo{volume}{49}}, \bibinfo{pages}{14211} (\bibinfo{year}{1994}).

\bibitem[{\citenamefont{A~Ferretti and Felice}(2007)}]{Ferretti07}
\bibinfo{author}{\bibfnamefont{B.~B.} \bibnamefont{A~Ferretti},
  \bibfnamefont{A~Calzolari}} \bibnamefont{and}
  \bibinfo{author}{\bibfnamefont{R.~D.} \bibnamefont{Felice}},
  \bibinfo{journal}{Journal of Physics: Condensed Matter}
  \textbf{\bibinfo{volume}{19}}, \bibinfo{pages}{036215 (16pp)}
  (\bibinfo{year}{2007}).

\bibitem[{\citenamefont{Mostofi et~al.}(2007 (in press))\citenamefont{Mostofi,
  Yates, Lee, Souza, Vanderbilt, and Marzari}}]{mostofi07}
\bibinfo{author}{\bibfnamefont{A.~A.} \bibnamefont{Mostofi}},
  \bibinfo{author}{\bibfnamefont{J.~R.} \bibnamefont{Yates}},
  \bibinfo{author}{\bibfnamefont{Y.-S.} \bibnamefont{Lee}},
  \bibinfo{author}{\bibfnamefont{I.}~\bibnamefont{Souza}},
  \bibinfo{author}{\bibfnamefont{D.}~\bibnamefont{Vanderbilt}},
  \bibnamefont{and} \bibinfo{author}{\bibfnamefont{N.}~\bibnamefont{Marzari}},
  \bibinfo{journal}{Comput. Phys. Commun.}  (\bibinfo{year}{2007 (in press)}).

\end{thebibliography}
\end{document}